\RequirePackage{lineno}
\documentclass[aps,prl,showpacs,amsfonts,notitlepage,amssymb,amsmath,nofootinbib,superscriptaddress,longbibliography,twocolumn]{revtex4-1}
\usepackage[utf8]{inputenc}
\usepackage[english]{babel}
\usepackage[titletoc,toc,title]{appendix}
\usepackage{amsmath}
\usepackage{amsfonts}
\usepackage{amssymb}
\usepackage[colorlinks=true,linkcolor=blue,citecolor=blue,urlcolor=blue]{hyperref}
\usepackage{graphicx}
\usepackage{enumerate}
\usepackage{csquotes}
\usepackage[caption=false]{subfig}
\usepackage{dsfont}
\usepackage{color}
\usepackage{bbold}
\usepackage{mathtools}
\usepackage{tensor}
\usepackage[export]{adjustbox}
\usepackage{hyperref}

\begin{document}

\title{Scattering of co-current surface waves on an analogue black hole}
\date{\today}

\author{L\'{e}o-Paul Euv\'{e}}
\affiliation{Laboratoire de Physique et M\'{e}canique des Milieux H\'{e}t\'{e}rog\`{e}nes (UMR 7636), ESPCI, Universit\'{e} PSL, CNRS, Sorbonne Universit\'{e}, Univ. Paris Diderot, 10 rue Vauquelin, 75321 Paris Cedex 05, France}
\author{Scott Robertson}
\affiliation{Universit\'e Paris-Saclay, CNRS/IN2P3, IJCLab, 91405 Orsay, France}
\author{Nicolas James}
\affiliation{Laboratoire de Math\'{e}matiques et Applications (UMR 7348), CNRS--Universit\'{e} de Poitiers, 11 Boulevard Marie et Pierre Curie--T\'{e}l\'{e}port 2--BP 30179, 86962 Futuroscope Chasseneuil Cedex, France}
\author{Alessandro Fabbri}
\affiliation{Departamento de F\'{i}sica Te\'{o}rica and IFIC, Centro Mixto Universidad de Valencia--CSIC, C. Dr. Moliner 50, 46100 Burjassot, Spain}
\affiliation{Centro Fermi - Museo Storico della Fisica e Centro Studi e Ricerche Enrico Fermi, Piazza del Viminale 1, 00184 Roma, Italy}
\affiliation{Universit\'e Paris-Saclay, CNRS/IN2P3, IJCLab, 91405 Orsay, France}
\author{Germain Rousseaux}
\affiliation{Institut Pprime (UPR 3346), CNRS--Universit\'{e} de Poitiers--ISAE ENSMA, 11 Boulevard Marie et Pierre Curie--T\'{e}l\'{e}port 2--BP 30179, 86962 Futuroscope Chasseneuil Cedex, France}

\begin{abstract}

We report on what is to our knowledge the first scattering experiment of surface waves on an accelerating transcritical flow, which in the Analogue Gravity context is described by an effective spacetime with a black-hole horizon.  
This spacetime has been probed by an incident co-current wave, which partially scatters into an outgoing counter-current wave on each side of the horizon.  
The measured scattering amplitudes are compatible with the predictions of the hydrodynamical theory, where the kinematical description in terms of the effective metric is exact.

\end{abstract}

\maketitle

Recent years have witnessed an explosion of interest in Analogue Gravity~\cite{LivingReview,Barcelo-2018}. 
This newly emergent field builds on Unruh's insight of 1981~\cite{Unruh-1981} that, using the 
mathematical equivalence between wave propagation in curved spacetimes on the one hand 
and in condensed matter systems on the other, 
one can realize 
an analogue black hole in the laboratory 
and test Hawking's prediction that black holes emit radiation~\cite{Hawking-1974,Hawking-1975}. 
In this context, an 
analogue black hole is engendered 
by an accelerating flow which 
becomes supersonic. 
The point where the flow crosses 
the speed of sound is the (acoustic) horizon for sound waves, in analogy with the (event) horizon for light at 
a gravitational black hole.
Because wave propagation close to the horizon is equivalent in the two cases, the Hawking effect, which is derived from purely kinematical considerations, should be present also in the analogue system.
This prediction 
has been the main driving force behind the Analogue Gravity program since its inception, with 
experimental studies in a wide range of physical systems, such as 
nonlinear optics~\cite{Philbin-et-al-2008,Belgiorno-et-al-2010,Ciret-et-al-2016,Drori-et-al-2019},  
exciton-polaritons~\cite{Nguyen-et-al-2015} and Bose-Einstein condensates~\cite{Steinhauer-2014,Steinhauer-2016,Steinhauer-2019}. 

Surface waves on fluids provide another example of an analogue system~\cite{Schuetzhold-Unruh-2002,Rousseaux-BASICS-2013}. 
Consider a rectangular flume oriented along the $x$-direction, 
containing a water flow which is independent of the transverse coordinate $y$. 
In the shallow-water limit $k h \ll 1$ (
$k$ being 
the wave vector and $h$ 
the 
depth), 
the free-surface deformation $\delta h$ behaves like the canonical momentum of a 
minimally coupled 
massless scalar field propagating in the $(2+1)$-dimensional spacetime metric
\begin{linenomath}\begin{equation}
\mathrm{d}s^{2} = c^{2} \left[ c^{2} \, \mathrm{d}t^{2} - \left(\mathrm{d}x - v \, \mathrm{d}t \right)^{2} - \mathrm{d}y^{2} \right] \,.
\label{eq:PGmetric}
\end{equation}\end{linenomath}
The metric~(\ref{eq:PGmetric}) describes a ``spacetime fluid'' flowing in the $x$-direction with velocity $v(x)$ and variable ``speed of light'' $c(x)$. 
The $x$-oriented 
null geodesics of the metric~(\ref{eq:PGmetric}) 
are simply described by ${\rm d}x/{\rm d}t = v \pm c$.  
For waves, this equation gives the motion of characteristics, and 
is equivalent to the following (linear) dispersion relation whenever $v$ and $c$ are constant:
\begin{linenomath}\begin{equation}
\Omega = \omega - v k = \pm c k \,,
\label{eq:Doppler}
\end{equation}\end{linenomath}
where $\Omega$ is the ``co-moving'' frequency measured in the rest frame of the fluid, while $\omega$ is the Doppler-shifted frequency measured in the lab frame.
We assume the flow is from left to right, i.e. $v > 0$, so that the plus and minus signs in Eq.~(\ref{eq:Doppler}) respectively define 
waves engaged in {\it co-} and {\it counter-current} propagation. 
In particular, the total velocity of counter-current waves is $v-c$, which can take either sign.  
A {\it transcritical} flow has $v-c = 0$ somewhere. 
This point is the (analogue) horizon, marking the boundary between {\it subcritical} ($v-c < 0$) and {\it supercritical} ($v-c > 0$) regions.
Accelerating flows which pass 
from subcritical to supercritical 
have specifically an analogue {\it black-hole} horizon, 
with counter-current waves outgoing on either side.

Matters are complicated by the prefactor of $c^{2}$ in Eq.~(\ref{eq:PGmetric}), which can be thought of as a conformal factor relating the effective metric to the simpler one 
in square brackets.  In an appropriate coordinate system, 
this term generates an effective potential~\cite{Anderson-et-al-2013,Anderson-et-al-2014} which 
scatters co- into counter-current waves and {\it vice versa}.  The two outgoing counter-current waves 
can thus be excited by a co-current wave sent in by a wave maker.

There have been a number of recent experiments aimed at realizing and probing analogue spacetimes of the form~(\ref{eq:PGmetric}) seen by surface waves~\cite{Rousseaux-et-al-2008,Rousseaux-et-al-2010,Weinfurtner-et-al-2011,Euve-et-al-2015,Euve-et-al-2016,Torres-et-al-2017,Euve-Rousseaux-2017}.
However, these have tended to remain far from the ideal case originally envisaged by Unruh.
Most importantly, in none of these experiments was a transcritical flow achieved, so that the corresponding analogue spacetimes did not contain horizons~\footnote{A rotational vortex, as an analogue of a rotating $(2+1)$-dimensional spacetime with an ergoregion, has been realized in~\cite{Torres-et-al-2017}, but no mention is made of the possible presence of a horizon. 
Note that dispersive models admit what may be termed ``group velocity horizons''~\cite{LivingReview}, i.e., turning points, but we use the term ``horizon'' in the more restrictive 
sense associated with the effective metric.}.
In such cases, the nontrivial scattering processes analogous to those involved in the Hawking effect can exist only thanks to the presence of short-wave dispersion, which requires the singling out of a preferred frame and thus breaks the analogue of Lorentz invariance. 
The relevant scattering amplitudes cannot be derived from the spacetime description; in particular, those pertaining to the analogue of the Hawking process do not generally follow the Planck law so central to Hawking's thermal prediction~\cite{Michel-Parentani-2014,Robertson-et-al-2016}.  
Furthermore, all experiments in 1D settings have focused on decelerating flows, which in the transcritical case 
would yield analogue white-hole horizons. 
These are convenient for scattering experiments as they convert incident hydrodynamical (long-wavelength) waves, which are easy to excite, into outgoing dispersive (short-wavelength) waves, which are easy to measure.
However, this means that the kinematics of the outgoing modes are not describable using the effective metric formalism alone.
Things are further complicated by the ubiquity of undulations (i.e., zero-frequency solutions with finite wave vector) on the downstream sides of decelerating flows~\cite{Coutant-Parentani-2014,Michel-et-al-2018}.

\section{Experimental setup}

\begin{figure*}
\includegraphics[width=0.55\textwidth]{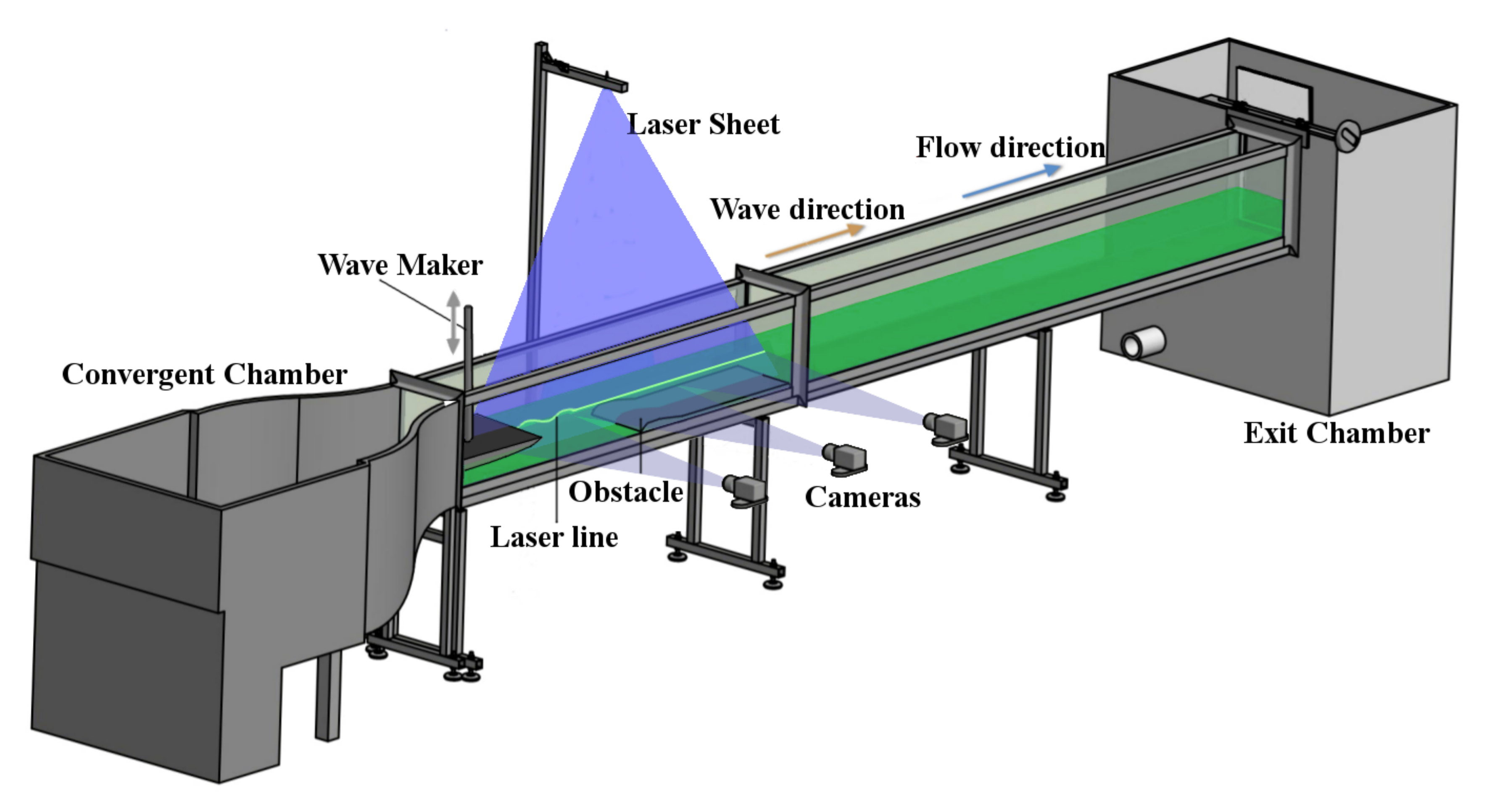} \, \includegraphics[width=0.4\textwidth]{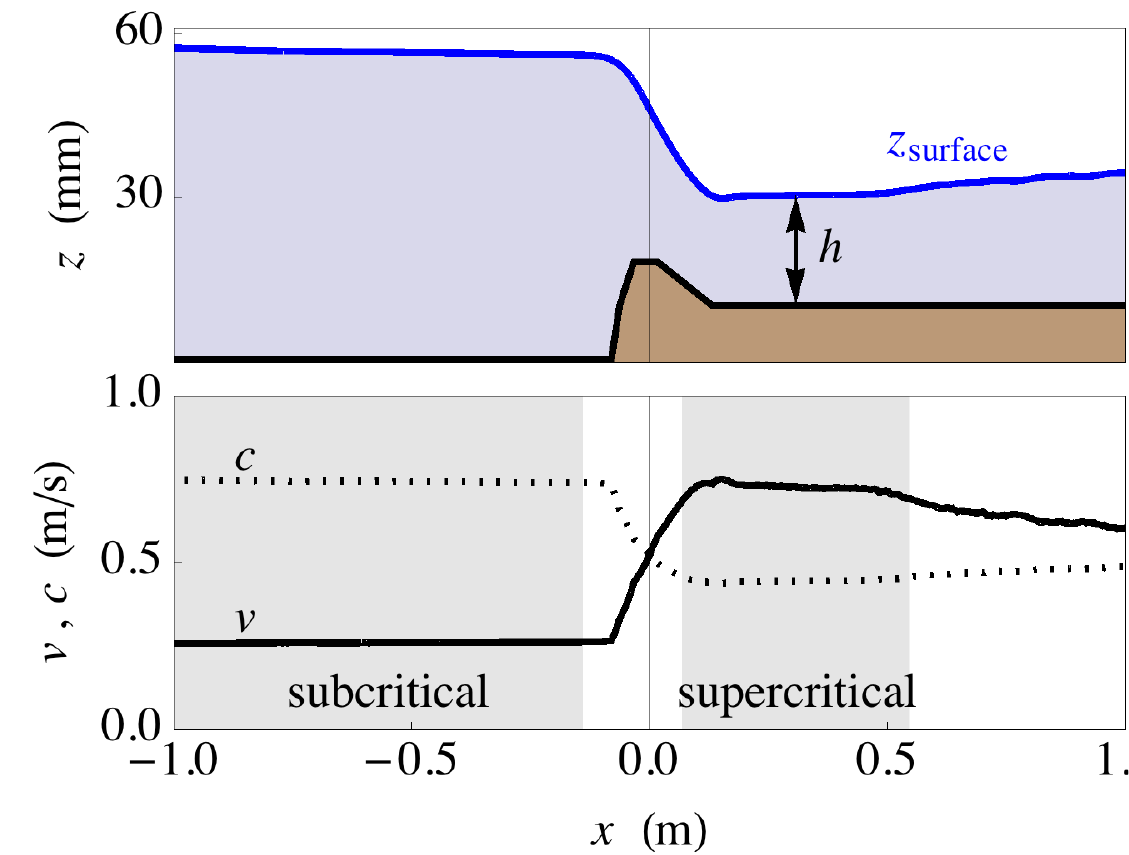} \\ 
\includegraphics[width=0.95\textwidth]{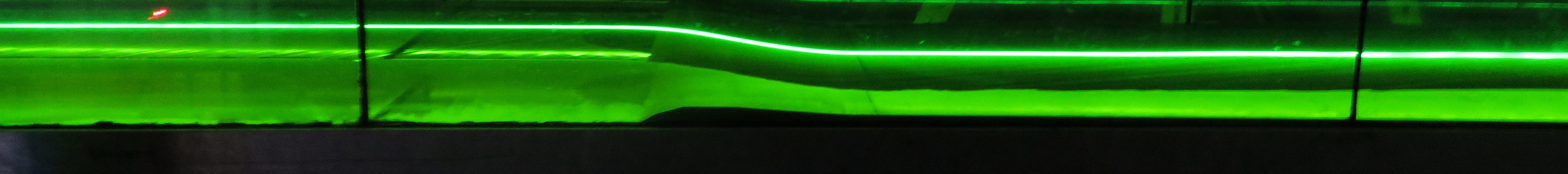}
\caption{Experimental setup.  On the left is an illustration of the flume, showing the flow direction and the position of the wave maker.  On the right are 
shown: upper) the obstacle on the bottom of the flume and the stationary profile of the surface, found by averaging $z_{\rm surface}(t,x)$ over time; lower)
the corresponding profiles of $v$ and $c$ (according to Eqs.~(\ref{eq:v_and_c})), with
the near-homogeneous regions (where the plane wave decomposition is performed) shaded.
The bottom 
panel shows a photo of the flume, where one sees clearly the variation of the water depth and the laser illuminating the midline.  
The distance between the black vertical bars is $1\,\mathrm{m}$. 
\label{fig:setup}}
\end{figure*}

We have realized, at Institut Pprime of the University of Poitiers, a transcritical flow in a flume analogous to a black-hole spacetime, and performed a scattering experiment on this flow that can be described in terms of the effective metric.
The experimental setup and the stationary 
flow profile are illustrated in Figure~\ref{fig:setup}.
A flume of length $7\,{\rm m}$ and width $39\,{\rm cm}$ is used, and on 
the bottom of the flume is placed a dedicated 
obstacle with $x$-dependent height, 
designed in accordance with well-known principles of hydraulics (set out pedagogically by Walker in~\cite{WalkerSciAm}). 
The 
current 
is maintained 
by a PCM Moineau pump.
The convergent chamber, 
containing a honeycomb structure, removes practically all boundary-layer effects and macro-vortices, producing a flow velocity profile that (in the upstream region) 
is approximately constant in both the vertical and tranverse directions. 
At a given flow rate, 
which is measured by a Promag 55S flow meter to within an error of $0.5 \%$,
the stationary flow that results has a particular mean depth profile $h(x)$.
The presence of surface waves 
causes the instantaneous depth to vary in time around this mean.

Measurements were made of the instantaneous vertical position $z_{\rm surface}(t,x)$ 
of the water surface along the midline of the flume.
A laser sheet, produced by a laser diode (at wavelength $473\,\mathrm{nm}$ and 
power 
$100\,\mathrm{mW}$) and a Powell lens, was shone through the water, to which had been added fluorescein dye~\cite{Weinfurtner-et-al-2011,Euve-et-al-2016}. 
At a suitable concentration ($\sim 20 \, {\rm g}/{\rm m}^{3}$), the dye absorbs the blue laser light and emits green light near the surface (up to a depth of $\sim 5 \, \mathrm{mm}$). 
The laser line produced (see 
Fig.~\ref{fig:setup}) was recorded by three cameras at regular intervals of $\Delta t \sim 0.1\,{\rm s}$, for a total duration of $800\,{\rm s}$.
A subpixel detection algorithm yielded the surface height 
at each time 
at a series of positions separated by the 
constant spacing 
$\Delta x = 3.4\,{\rm mm}$. 
From this was subtracted the known height profile of the obstacle in order to get the instantaneous water depth $h(t,x)$. 
The mean depth $h(x)$ 
is then found by averaging over time.
Assuming the fluid to be irrotational and the surface quasi-flat (i.e. $\left| \partial_{x}z_{\rm surface} \right| \ll 1$), 
the flow velocity and 
wave speed are related to the mean depth by
\begin{linenomath}\begin{equation}
v(x) = \frac{q}{h(x)} \,, \qquad c(x) = \sqrt{g h(x)} \,,
\label{eq:v_and_c}
\end{equation}\end{linenomath}
where $q$ is the (constant) flow rate per unit width (measured at $147.4 \pm 0.7 \,{\rm cm}^{2}/{\rm s}$) 
and $g = 9.8\,\mathrm{m/s^{2}}$ is the acceleration due to gravity.
As seen in the lower right panel of Fig.~\ref{fig:setup}, the realized flow passes from subcritical to supercritical, thus engendering an analogue black-hole horizon at the point where $v=c$. 

\begin{figure}
\includegraphics[width=0.95\columnwidth]{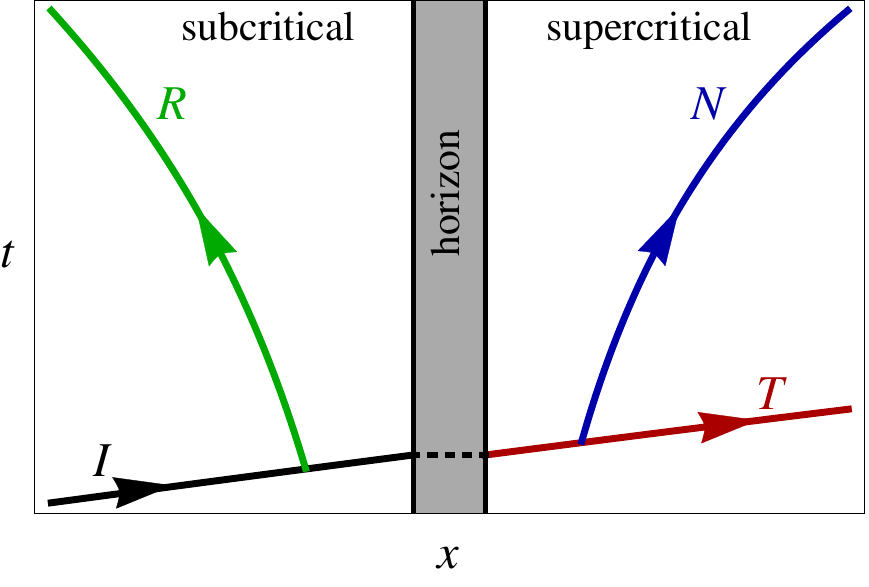}
\caption{Space-time diagram of the scattering process, showing 
the wave characteristics involved in the 
scattering of an incident co-current wave ({\it I}, black) into a transmitted co-current wave ({\it T}, red), a reflected counter-current wave ({\it R}, green) and a transmitted counter-current wave ({\it N}, blue). 
The flow is 
from left to right, passing 
from subcritical to supercritical.  
\label{fig:v-scattering}}
\end{figure}

Surface waves were stimulated by a plunging-type wave maker placed upstream from the obstacle. 
We performed a series of experimental runs for a range of frequencies at 
three different wave maker amplitudes: 
$A_{\rm wm} = 0.25\,{\rm mm}$, for which $f \in \left[ 0.55\,,\, 1.2 \right] \, {\rm Hz}$; $A_{\rm wm} = 0.5\,{\rm mm}$, for which $f \in \left[ 0.2 \,,\, 1.0 \right] \, {\rm Hz}$; and $A_{\rm wm} = 1.0\,{\rm mm}$, for which $f \in \left[ 0.2\,,\, 0.5 \right] \, {\rm Hz}$. 
The elongated shape of the wave maker allows the displacement of a large volume of fluid, producing long waves of small amplitude which 
have a negligible effect on the mean flow (we estimate the relative change in the Froude number $v/c$ due to the passage of the wave to be at most $2\%$). 
The co-current waves thus 
excited were incident on the horizon; 
this is analogous to sending matter 
in towards a black hole from far away.
The scattering process is illustrated in Figure~\ref{fig:v-scattering}. 
Much of the wave was simply transmitted (with amplitude $T$), while scattering at the effective potential also produced a reflected wave (with amplitude $R$) and a 
counter-current wave in the supercritical region (with amplitude $N$).

\section{Main results}

Focusing 
on the near-homogeneous regions 
(shaded gray 
in the 
lower 
right panel of Fig.~\ref{fig:setup}), 
we assumed that the waveform in each 
is 
a sum of two plane waves, and were 
thus able to extract best-fit values for the amplitudes and wave vectors -- see the Supplemental Material (SM)~\cite{SM} for details. 
The subcritical region staying 
relatively flat over a considerable distance, we could 
take a longer near-homogeneous region there, and we thus 
expect greater accuracy of the results pertaining 
to this side.

\begin{figure}
\includegraphics[width=0.95\columnwidth]{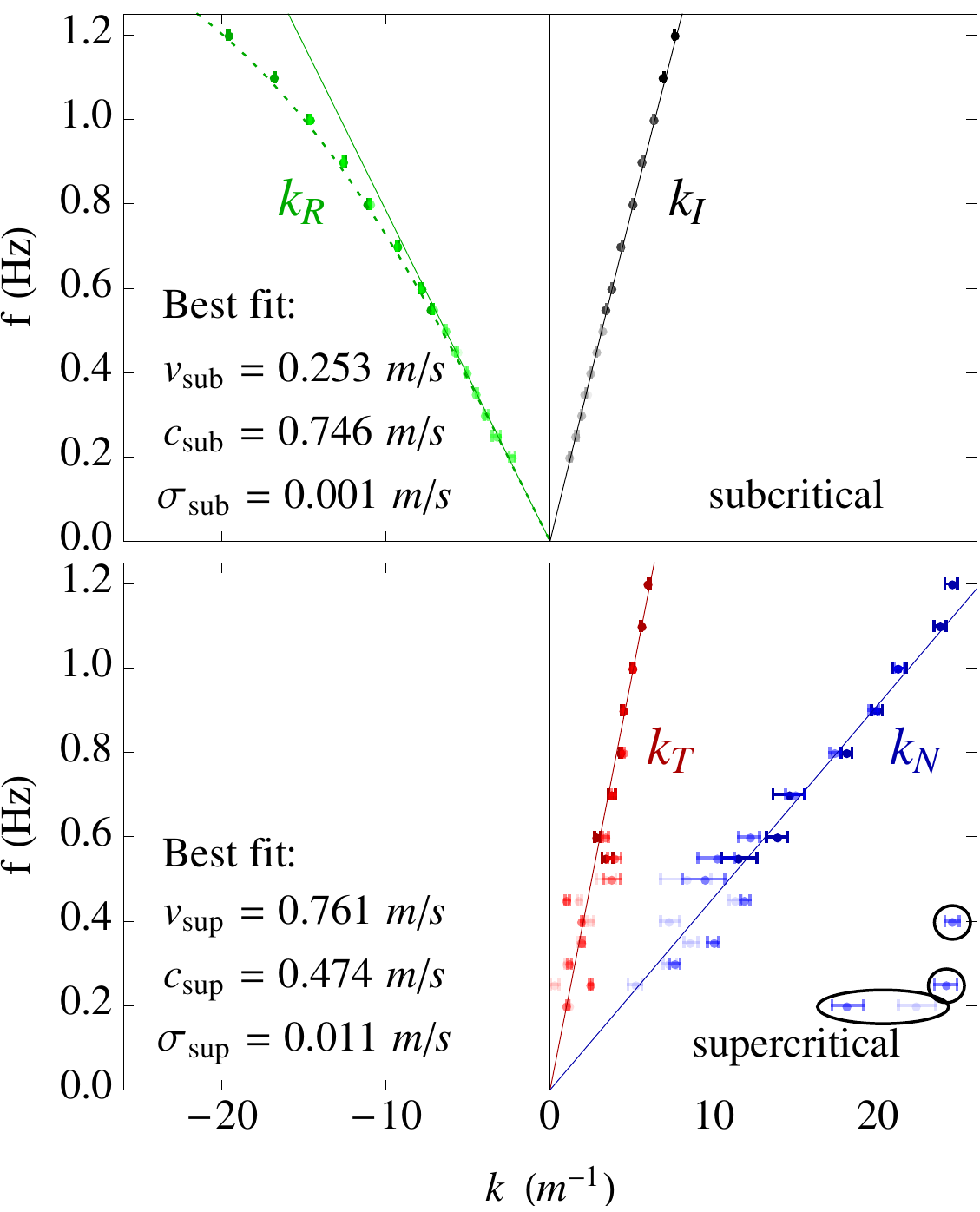} 
\caption{Measured wave vectors. 
The various colors correspond to the 
modes of Fig.~\ref{fig:v-scattering}, while the three different shadings of data points correspond to different wave maker amplitudes: $A_{\rm wm} = 0.25 \, {\rm mm}$ (dark), $0.5 \, {\rm mm}$ (medium) and $1 \, {\rm mm}$ (light).  
Circled points are outliers.
The solid curves show theoretical dispersion relations fitted to the linear behavior of Eq.~(\ref{eq:Doppler}) for $k h \ll 1$.  
The fitted values of $v$ and $c$ in each region, as well as their error $\sigma$, are indicated. 
The dotted line in the upper panel shows the full theoretical dispersion relation (see footnote~\ref{fn:disp_rel_tanh}) with $h_{\rm sub}$ given by $c_{\rm sub}$ via Eqs.~(\ref{eq:v_and_c}). 
\label{fig:k-fitted}}
\end{figure}

{\it Dispersion relation.} 
In Figure~\ref{fig:k-fitted} are shown the best-fit values of the wave vectors in each of the near-homogeneous regions. 
Immediately, we see that (with the exception of three 
outliers on the bottom right of the lower panel) 
each region 
shows two 
branches of the dispersion relation with clear signs
of the derivative ${\rm d}f/{\rm d}k$ (where $f = \omega/2\pi$), this being 
the sign of the group velocity of the corresponding wave.  We thus conclude 
that we have one ingoing and one outgoing wave on the subcritical side, while on the supercritical side we instead have two outgoing waves.  
This 
is indicative of the presence of an analogue black-hole horizon: it shows that $v-c$ changes sign when moving from one near-homogeneous region to the other.

Also shown are fits to the theoretical dispersion relations.  As already mentioned, these take the linear form of Eq.~(\ref{eq:Doppler}) in the shallow-water regime $k h \ll 1$, so we may use the lower range of $k$ to find the best-fit values of $v$ and $c$ in each region.  These are given explicitly in Fig.~\ref{fig:k-fitted}; $v$ is about $3\%$ smaller than predicted by Eqs.~(\ref{eq:v_and_c}) (see the discussion below concerning the flow rate $q$), while $c$ agrees with the prediction to within $0.1\%$..  
Moreover, since we expect the flow to be uniform in the subcritical region, we may use the best-fit values of $v$ and $c$ to find the corresponding values of $q$ and $h$ through Eqs.~(\ref{eq:v_and_c}).  
Not only is the extracted water depth there, $h_{\rm sub} = 56.7 \pm 0.1\,{\rm mm}$, consistent with the profile of Fig.~\ref{fig:setup}, 
but it also agrees with the observed deviations from the linear dispersion relation~\footnote{These dispersive corrections to the phase velocity 
are expected to take the form $c(k) = \sqrt{g h} \times \sqrt{ {\rm tanh}\left(h k\right) / \left(h k\right)}$~\cite{Lamb}, and are thus 
governed by the water depth $h$. \label{fn:disp_rel_tanh}}. 
The fitted value of the flow rate is $q_{\rm fit} = 143.6 \pm 0.3 \, {\rm cm}^{2}/{\rm s}$, which differs quite significantly from the measured value of $147.4 \pm 0.7 \,{\rm cm}^{2}/{\rm s}$.  
This is most likely due to the non-uniformity of the flow speed in the vertical direction.  Indeed, numerical simulations indicate that $v$ tends to decrease slightly near the surface, which would explain why $q_{\rm fit}$ is smaller than the measured value (see SM~\cite{SM}).

On the supercritical side, 
the extracted wave vectors are considerably less accurate, having larger error bars and appearing to oscillate around 
the expected linear curves. 
While this is partly due to the reduction of the window size with respect to that in the subcritical region, 
it should be noted that there is a considerable degree of noise in the supercritical region. 
Even so, only four 
data points (which have been circled in Fig.~\ref{fig:k-fitted}) are clear outliers.
$v$ and $c$ 
are found to be about $5$ and $7 \%$ larger, 
respectively, than those predicted by Eqs.~(\ref{eq:v_and_c}). 
This discrepancy is likely due to the non-uniformity of the flow after passing over the obstacle, 
and is reminiscent of a similar observation made in the downstream region in Ref.~\cite{Euve-et-al-2015}.

\begin{figure}
\includegraphics[width=0.95\columnwidth]{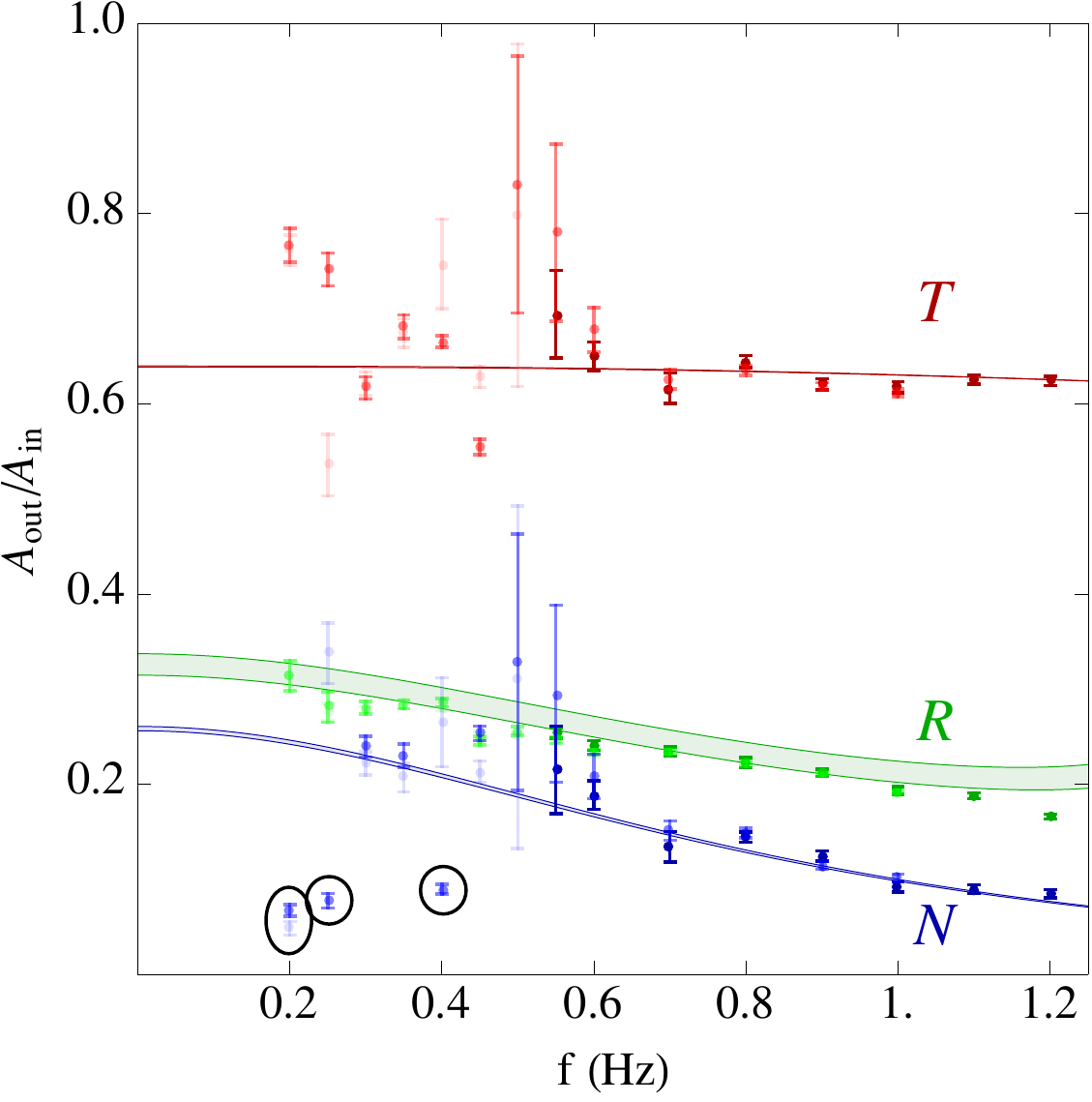} 
\caption{Measured scattering amplitudes.  
The colors and shadings correspond 
to the same outgoing branches and wave maker amplitudes as in Fig.~\ref{fig:k-fitted}.  
The colored lines are theoretical predictions, and have finite widths corresponding to the uncertainty in $h_{\rm sub}$.
The amplitudes of the 
outliers circled in Fig.~\ref{fig:k-fitted} are also circled here.
\label{fig:S-coeffs}}
\end{figure}

{\it Scattering coefficients.} 
In Figure~\ref{fig:S-coeffs} are shown the three outgoing wave amplitudes in units of the incident wave amplitude.
Also shown are predictions extracted from the wave equation associated with the metric~(\ref{eq:PGmetric})~\footnote{To improve the agreement between the theoretical prediction and the experimental data we have used refined versions of Eqs.~(\ref{eq:v_and_c}) (see Eqs.~(21) of SM~\cite{SM}).  
We have also included dispersive corrections to the conversion factors relating normalized and ``bare'' amplitudes (see Eqs.~(15) of SM~\cite{SM}).}. 
As expected, the most accurate of the scattering amplitudes (in terms of the smallness of the error bars) is $R$, this having been extracted purely from the subcritical region. 
The amplitudes relating to the supercritical region are less accurately determined, with much larger error bars than $R$.
This is due both to the smaller available window size where the flow is approximately homogeneous, and also to the presence of a significant amount of noise, which seems to be related to the occurrence of side wakes (see SM~\cite{SM}).
However, the extracted scattering amplitudes 
still compare favorably with the hydrodynamical theory.

In summary, 
we have extracted the dispersion relations and scattering amplitudes pertaining to the interaction of an incident co-current surface wave with an analogue black hole.
The dispersion relations show a total of three outgoing modes, implying the presence of an analogue horizon. 
Except for the high-frequency sector of the $R$-branch, the modes relevant to the scattering process are in the non-dispersive regime where the effective metric description is valid, and 
the measured scattering amplitudes are compatible with its predictions.

Following in the spirit of~\cite{Weinfurtner-et-al-2011,Euve-et-al-2016}, a 
future goal of this type of (classical) 
analogue model is the observation of (the stimulated version of) the Hawking process, whereby the squared norm of the relevant scattering coefficient 
takes the form of a thermal spectrum.
The absence of transcriticality in previous experiments prevents the direct verification of the Hawking-Unruh thermal prediction, which can only be applied when a horizon is present.
Our setup makes this possible in two ways.
Firstly, one could consider a stationary background like that used in this paper.
This can only be achieved by exploiting dispersion and sending in a short-wavelength mode~\footnote{There are two possibilities for the incident dispersive mode: either a short-wavelength gravity wave incident from the subcritical side, or a capillary wave travelling at ``superluminal'' speed with respect to the fluid and incident from the supercritical side.  The conversion of 
capillary waves to 
long-wavelength gravity waves was demonstrated in the wormhole travel experiment~\cite{Euve-Rousseaux-2017}, 
despite 
a high viscous damping rate for the capillary waves~\cite{Robertson-Rousseaux}.}, 
but these are difficult to excite in practice.
The second possibility is to consider a time-dependent background in which an initially subcritical flow develops a supercritical region.
In this context one can study, in particular, the scattering of initial long-wavelength countercurrent modes into final long-wavelength countercurrent modes (i.e., those labeled as $R$ and $N$ in the present work), thus avoiding the need to send in a dispersive wave.
This case 
is reminiscent of 
that originally studied by Hawking~\cite{Hawking-1974,Hawking-1975}, of a star collapsing to form a black hole. 

\subsection*{Acknowledgments}

We would like to thank Romain Bellanger, Laurent Dupuis and Jean-Marc Mougenot for their help concerning the experimental aspects of this work.
We also thank Roberto Balbinot, Ulf Leonhardt, and particularly Florent Michel, Renaud Parentani and Thomas Philbin for their careful reading of and constructive comments on a previous version of the manuscript. 
This work was supported by the French National Research Agency through Grant No. ANR-15-CE30-0017-04 associated with the project HARALAB,
by the University of Poitiers (ACI UP on Wave-Current Interactions 2013--2014), 
by the Interdisciplinary Mission of CNRS which funded the linear motor of the wave maker in 2013, 
and by the University of Tours in a joint grant with the University of Poitiers (ARC Poitiers-Tours 2014--2015).
It benefited from the support of the project OFHYS of the CNRS 80 Pprime initiative in 2019.
It was partially supported by the Spanish Mineco grants FIS2014-57387-C3-1-P and FIS2017-84440-C2-1-P, the Generalitat Valenciana project SEJI/2017/042 and the Severo Ochoa Excellence Center Project SEV-2014-0398.


\nocite{Unruh2013}
\nocite{Corley-Jacobson-1996}
\nocite{Macher-Parentani-BW}
\nocite{DeWitt-1975}
\nocite{Baird-1970}
\nocite{Massar-Parentani-1998}
\nocite{Anderson-Fabbri-Balbinot-2015}
\nocite{IBM}
\nocite{LST}

\bibliography{biblio}


\begin{appendices}

\section{Supplemental material}

\subsection{Theoretical background}

In this section we provide more theoretical details on 
the effective metric description of the scattering of surface waves,
with particular emphasis on the scattering process examined in the main paper (referred to here as the Letter).

\subsubsection{Wave equation}
 
In the shallow-water limit $h/\lambda \ll 1$ (where $h$ is the water depth and $\lambda$ the wavelength), 
and assuming $\left|\partial_{x}z_{\rm surface}\right| \ll 1$ everywhere (where $z_{\rm surface} = h_{\rm obs} + h$ is the vertical position 
of the free surface, $h_{\rm obs}$ being the height of the obstacle), 
longitudinal surface waves on an inviscid, irrotational, two-dimensional flow obey the following wave equation~\cite{Schuetzhold-Unruh-2002,Unruh2013,Coutant-Parentani-2014}:
\begin{linenomath}\begin{equation}
\left[ \left( \partial_{t} + \partial_{x} v \right) \left( \partial_{t} + v \partial_{x} \right) - \partial_{x} c^{2} \partial_{x} \right] \phi = 0 \,.
\label{eq:wave_eqn}
\end{equation}\end{linenomath}
The scalar field $\phi$ is a perturbation of the velocity potential, i.e., $\Phi = \Phi_{0} + \phi$ where $\vec{v} = -\nabla\Phi$.  The free-surface deformation is given by $\delta h = -\frac{1}{g}\left(\partial_{t}+v\partial_{x}\right) \phi$ (where $g = 9.8\,{\rm m}/{\rm s}^{2}$ is the acceleration due to gravity).  
The position-dependent parameters $v$ and $c$ are 
determined by the background flow; specifically, $v$ is the longitudinal component of the local flow velocity 
and $c$ is the local speed of surface waves with respect to the fluid.
Equation~(\ref{eq:wave_eqn}) is precisely the Klein-Gordon wave equation in curved spacetime, $\Box \phi = 0$ (see~\cite{LivingReview} for details), for the longitudinal modes of a massless scalar field propagating in the $(2+1)$-dimensional spacetime metric~\footnote{We have neglected an overall factor which would make Eq.~(\ref{eq:PGmetric}) dimensionally correct; this is unimportant for our present purposes.  It should also be noted that, if one were to include two transverse coordinates, $y$ and $z$, then the overall factor of $c^{2}$ would have to be replaced by $c$ in order to reproduce Eq.~(\ref{eq:wave_eqn}), as given in Eq.~(A32) of~\cite{Coutant-Parentani-2014} with $\rho(x) \propto c^{2}(x)$.  We present the $(2+1)$-dimensional form here as it seems a more natural choice for surface waves.}
\begin{linenomath}\begin{equation}
\mathrm{d}s^{2} = c^{2} \left[ c^{2} \, \mathrm{d}t^{2} - \left(\mathrm{d}x - v \, \mathrm{d}t \right)^{2} - \mathrm{d}y^{2} \right] \,,
\label{eq:PGmetric}
\end{equation}\end{linenomath}
already given in Eq.~(1) of the Letter.
As mentioned there, the
characteristics of 
wave equation~(\ref{eq:wave_eqn}) are precisely the $x$-directed null geodesics of the metric~(\ref{eq:PGmetric}), and are simply described by ${\rm d}x/{\rm d}t = v \pm c$, which in the wave picture is equivalent to the dispersion relation
\begin{linenomath}\begin{equation}
\Omega = \omega - v k = \pm c k
\label{eq:Doppler}
\end{equation}\end{linenomath}
whenever $v$ and $c$ are constant.
While $\omega$ is just the Doppler shifted frequency due to the motion of the fluid, 
stationarity in the frame of the flume ensures that $\omega$ (rather than $\Omega$) is conserved by any linear scattering process. 
Assuming for definiteness that $v>0$, dispersion relation~(\ref{eq:Doppler}) defines {\it co-current} and {\it counter-current} waves (respectively, those taking the $+$ and $-$ sign on the right-hand side of Eq.~(\ref{eq:Doppler})), and a horizon where $v-c=0$ splits the counter-current waves into two distinct branches localized on either side of the horizon.

\subsubsection{Conservation laws} 

The conserved energy 
associated with Eq.~(\ref{eq:wave_eqn}) 
is (up to an overall prefactor)
\begin{linenomath}\begin{equation}
E = \int \mathrm{d}x \left\{ \left|\partial_{t}\phi\right|^{2} + \left(c^{2}-v^{2}\right) \left|\partial_{x}\phi\right|^{2} \right\} \,.
\label{eq:Hamiltonian}
\end{equation}\end{linenomath}
From this expression, 
it is readily apparent that wherever the flow becomes supercritical, it is possible for the integrand (i.e., the energy density) to be negative. 
Treating $v$ and $c$ as constant and directly substituting $\phi = A\,{\rm exp}\left(ikx-i\omega t\right)$, with $\omega$ and $k$ satisfying  
dispersion relation~(\ref{eq:Doppler}), 
we find that the energy density of such a plane wave is (proportional to) $\omega \Omega \left|A\right|^{2}$. 
The signs of $\omega$ and $\Omega$ are opposite 
whenever the flow is strong enough to reverse the total velocity of the wave, which 
is precisely the case for counter-current waves in the supercritical region. 
These waves thus have negative energy with respect to the stationary background: their presence {\it reduces} the total energy of the system.
Crucially, they share their lab frequency $\omega$ with several waves of positive energy, yielding the possibility of {\it anomalous} scattering involving both positive- and negative-energy waves.

At fixed $\omega$, the wave form is stationary and infinitely extended, and the most useful conservation law is that for the {\it energy flux}, defined as the energy density times the group velocity $v_{g} = {\rm d}\omega/{\rm d}k = v \pm c$, where the $+$ ($-$) corresponds to co- (counter-) current waves.
It is straightforward to show that, for a plane wave, the energy flux is proportional to $\pm \omega^{2} c \left|A\right|^{2}$; therefore, co- (counter-) current waves always contribute positively (negatively) to the total energy flux.
The total energy flux must be constant in $x$; for, if it were not, the energy density would be changing somewhere, contradicting the stationarity of the wave form.
It is thus appropriate to consider normalized plane waves $\phi^{(N)}$ of fixed 
energy flux, so that a general plane wave $\phi$, 
and the corresponding free-surface deformation $\delta h = -\frac{1}{g}\left(\partial_{t}+v\partial_{x}\right)\phi$, 
can be written as
\begin{linenomath}\begin{alignat}{4}
\phi &=& \mathcal{A} \, \phi^{(N)} &=& \frac{\mathcal{A}}{\sqrt{\omega}} \, \frac{{\rm exp}\left(ikx - i\omega t\right)}{\sqrt{\omega \, c}} \,, \nonumber \\
\delta h &=& \mathcal{A} \, \delta h^{(N)} &=& \frac{\mathcal{A}}{\sqrt{\omega}} \, i \frac{\Omega}{g} \, \frac{{\rm exp}\left(ikx - i\omega t\right)}{\sqrt{\omega \, c}} \,.
\label{eq:normalization}
\end{alignat}\end{linenomath}
The contribution of this wave to the total energy flux is proportional to $\pm \left|\mathcal{A}\right|^{2}$.

Equations~(\ref{eq:normalization}) generalize readily to dispersive models~\cite{Corley-Jacobson-1996,Macher-Parentani-BW}:
\begin{linenomath}\begin{alignat}{4}
\phi &=& \mathcal{A} \, \phi^{(N)} &=& \frac{\mathcal{A}}{\sqrt{\omega}} \, \frac{{\rm exp}\left(ikx - i\omega t\right)}{\sqrt{\left|\Omega \, v_{g}\right|}} \,, \nonumber \\
\delta h &=& \mathcal{A} \, \delta h^{(N)} &=& \frac{\mathcal{A}}{\sqrt{\omega}} \, i \frac{\Omega}{g} \, \frac{{\rm exp}\left(ikx - i\omega t\right)}{\sqrt{\left|\Omega \, v_{g}\right|}} \,,
\label{eq:normalization_dispersive}
\end{alignat}\end{linenomath}
where $v_{g} = {\rm d}\omega/{\rm d}k$ is the group velocity.  In the non-dispersive limit, these reduce to Eqs.~(\ref{eq:normalization}).  Most important for our purposes is the factor of $\sqrt{\left|\Omega/v_{g}\right|}$ multiplying the plane waves of $\delta h$, since at fixed $\omega$ this is the part of the normalizing prefactor which varies from one mode to another.

\subsubsection{Scattering of co-current waves}

As mentioned in the Letter, the 
simple kinematical description given above is complicated by the prefactor $c^{2}$ in Eq.~(\ref{eq:PGmetric}). 
In the present context, the significance of this term stems from the fact that it is the transverse component of the metric, i.e., the coefficient of ${\rm d}y^{2}$ in Eq.~(\ref{eq:PGmetric}). Although we are interested in waves which do not depend on $y$, 
local inhomogeneities in this term 
can scatter co- into counter-current waves and {\it vice versa}, 
thus playing the role of an effective potential~\footnote{This phenomenon is well-known when considering the radial behavior of waves in the Schwarzschild metric.  There, the transverse (i.e., angular) component of the metric varies with $r$ (in fact, it is proportional to $r^2$), and this radial dependence yields an effective potential that causes infalling and outgoing waves to scatter into each other; see e.g. Ref.~\cite{DeWitt-1975}.}.

Our experiment aims to probe the structure of the spacetime induced by a transcritical flow through the scattering of incident co-current waves by the effective potential.
The scattering process is illustrated in the space-time diagram shown in Figure~2 of the Letter~\footnote{Note that this is {\it not} the scattering process relevant to the analogue Hawking effect, where the ``incident'' waves (defined at time $t \to -\infty$) emanate from the horizon, and where the characteristic thermal spectrum arises from an appropriate matching condition across the horizon~\cite{Unruh-1981}.}.
A co-current wave is incident from the subcritical side; this is analogous to sending matter in towards a black hole from far away.
Since the effective potential vanishes at the horizon, the entire process can be decomposed into two time-ordered scattering events, each localized to one side of the horizon.
First, the incident wave partially scatters at the effective potential outside the black hole, producing a (positive-energy) counter-current wave that propagates back towards the subcritical region.
This process is described by a reflection coefficient $\mathcal{R}$.
The unreflected part of the incident wave then crosses the horizon, and is partially scattered by the effective potential inside the black hole.
This time, a counter-current wave of negative energy is produced, and propagates in the same direction as the transmitted wave, i.e. falling further into the supercritical region.
Relating the outcome of this process to the incident wave requires two scattering coefficients, $\mathcal{N}$ and $\mathcal{T}$.
Imposing equality of the total energy flux on either side of the horizon, we find $1-\left|\mathcal{R}\right|^{2} = \left|\mathcal{T}\right|^{2}-\left|\mathcal{N}\right|^{2}$, or equivalently,
\begin{linenomath}\begin{equation}
\left| \mathcal{T} \right|^{2} + \left| \mathcal{R} \right|^{2} - \left| \mathcal{N} \right|^{2}  = 1 \,.
\label{eq:unitarity}
\end{equation}\end{linenomath}
This is the {\it unitarity relation}, an algebraic constraint on the scattering coefficients which encodes total energy conservation.  The fact that one of the scattering coefficients counts negatively towards this relation indicates that anomalous scattering is taking place.

\subsubsection{Calculation of scattering coefficients for a given flow profile}

In the non-dispersive model described by wave equation~(\ref{eq:wave_eqn}),
the normalized scattering amplitudes can be calculated directly by re-expressing~(\ref{eq:wave_eqn}) in terms of WKB-like solutions (in the spirit of~\cite{Baird-1970,Massar-Parentani-1998}). 
Let us 
define
\begin{equation}
W_{\omega}^{U}(x) = \frac{\mathrm{exp}\left(i\omega \, U(x)\right)}{\omega \, \sqrt{c(x)}} \,, \qquad
W_{\omega}^{V}(x) = \frac{\mathrm{exp}\left(i\omega \, V(x) \right)}{\omega \, \sqrt{c(x)}} \,,
\end{equation}
where
\begin{equation}
U(x) = \int^{x} \frac{dx^{\prime}}{v(x^{\prime})-c(x^{\prime})} \,, \qquad V(x) = \int^{x} \frac{dx^{\prime}}{v(x^{\prime})+c(x^{\prime})} \,.
\end{equation}
In homogeneous regions $W_{\omega}^{U}$ and $W_{\omega}^{V}$ describe the counter- and co-current waves, respectively.  Moreover, the factors of $\left(\omega\sqrt{c(x)}\right)^{-1}$ ensure 
that their energy flux is constant (cf. Eqs.~(\ref{eq:normalization})), 
so they are physically allowable solutions as far as energy conservation is concerned.  
However, they are not exact solutions of Eq.~(\ref{eq:wave_eqn}), because we have not yet accounted for the mixing of the the two modes by inhomogeneities in $c$.  To include this mixing, 
we introduce $x$-dependent amplitudes $\mathcal{A}_{\omega}^{U}$ and $\mathcal{A}_{\omega}^{V}$ such that
\begin{subequations}\begin{eqnarray}
\phi_{\omega}(x) & = & \mathcal{A}_{\omega}^{U}(x) \, W_{\omega}^{U}(x) + \mathcal{A}_{\omega}^{V}(x) \, W_{\omega}^{V}(x) \,, \label{eq:phi_amplitudes} \\
\partial_{x} \phi_{\omega}(x) & = & i \omega \, U^{\prime}(x) \, \mathcal{A}_{\omega}^{U}(x) \, W_{\omega}^{U}(x) \nonumber \\
&& \qquad \quad + i \omega \, V^{\prime}(x) \, \mathcal{A}_{\omega}^{V}(x) \, W_{\omega}^{V}(x) \,.
\end{eqnarray}\end{subequations}
Equation~(\ref{eq:wave_eqn}) can now be re-expressed as a pair of coupled first-order differential equations for the amplitudes:
\begin{equation}
\partial_{x} \left[ \begin{array}{c} \mathcal{A}_{\omega}^{U} \\ \mathcal{A}_{\omega}^{V} \end{array} \right]  = \frac{1}{2} \, \frac{\partial_{x}c}{c} \, \left[ \begin{array}{cc} 0 & e^{-i\omega(U(x)-V(x))} \\ e^{i\omega(U(x)-V(x))} & 0 \end{array} \right] \left[ \begin{array}{c} \mathcal{A}_{\omega}^{U} \\ \mathcal{A}_{\omega}^{V} \end{array} \right] \,.
\label{eq:wave_eqn_amplitudes}
\end{equation}
It is straightforward to show that Eq.~(\ref{eq:wave_eqn_amplitudes}) implies
\begin{equation}
\partial_{x} \left[ \left| \mathcal{A}_{\omega}^{V} \right|^{2} - \left| \mathcal{A}_{\omega}^{U} \right|^{2} \right] = 0 \,,
\end{equation}
which, given that $W_{\omega}^{U}$ and $W_{\omega}^{V}$ are already normalized, is exactly the expression for conservation of energy flux.  The waveform corresponding to the scattering of incident co-current waves is found by setting 
$\mathcal{A}_{\omega}^{U} = 0$ arbitrarily close to the horizon, i.e., by imposing the non-existence of waves emanating from the horizon.
We then integrate Eq.~(\ref{eq:wave_eqn_amplitudes}) out into the asymptotic regions -- this is done separately on the subcritical and supercritical sides of the flow, since $W_{\omega}^{U}$ must be separately defined on each side of the horizon -- and the appropriate normalization is achieved by setting the amplitude of the incident wave 
to $1$.
The amplitudes $\mathcal{A}_{\omega}^{V}$ and $\mathcal{A}_{\omega}^{U}$ in the homogeneous regions are then identified with $\mathcal{T}$, $\mathcal{R}$ and $\mathcal{N}$, satisfying Eq.~(\ref{eq:unitarity}).

As already stated, the free surface deformation $\delta h$ is related to $\phi$ via $\delta h = -\frac{1}{g} \left(\partial_{t}+v\partial_{x}\right)\phi$.  So, if $\phi$ is given by Eq.~(\ref{eq:phi_amplitudes}), $\delta h$ is given by
\begin{eqnarray}
\delta h_{\omega}(x) & = & \frac{i \omega}{g} \left[ \frac{c(x)}{c(x)-v(x)} \, \mathcal{A}_{\omega}^{U}(x) \, W_{\omega}^{U}(x) \right. \nonumber \\ 
&& \qquad \left. + \frac{c(x)}{c(x)+v(x)} \, \mathcal{A}_{\omega}^{V}(x) \, W_{\omega}^{V}(x) \right] \nonumber \\
& = & \frac{i}{g} \left[ \frac{\sqrt{c(x)}}{c(x)-v(x)} \, \mathcal{A}_{\omega}^{U}(x) \, e^{i \omega \, U(x)} \right. \nonumber \\
&& \qquad \left. + \frac{\sqrt{c(x)}}{c(x)+v(x)} \, \mathcal{A}_{\omega}^{V}(x) \, e^{i \omega \, V(x)} \right] \,.
\end{eqnarray}
This expression yields the necessary conversion factors between the normalized amplitudes and the ``bare'' amplitudes extracted directly from the data.  Using a cursive style for the normalized amplitudes (as in Eq.~(\ref{eq:unitarity})) and a non-cursive style for the ``bare'' amplitudes (as in the Letter), we have
\begin{eqnarray}
R &=& \frac{c_{\rm sub}+v_{\rm sub}}{c_{\rm sub}-v_{\rm sub}} \, \mathcal{R} \,, \nonumber \\ 
T &=& \sqrt{\frac{c_{\rm sup}}{c_{\rm sub}}} \, \frac{c_{\rm sub}+v_{\rm sub}}{c_{\rm sup}+v_{\rm sup}} \, \mathcal{T} \,, \nonumber \\ 
N &=& - \sqrt{\frac{c_{\rm sup}}{c_{\rm sub}}} \, \frac{c_{\rm sub}+v_{\rm sub}}{v_{\rm sup}-c_{\rm sup}} \, \mathcal{N} \,.
\label{eq:bare_v_normalized}
\end{eqnarray}
On the other hand, generalizing to the dispersive case using Eqs.~(\ref{eq:normalization_dispersive}), these become
\begin{eqnarray}
R &=& \sqrt{\left| \frac{\Omega_{R}}{\Omega_{\rm in}}\right| \left| \frac{v_{g, {\rm in}}}{v_{g, R}} \right|} \, \mathcal{R} \,, \nonumber \\
T &=& \sqrt{\left| \frac{\Omega_{T}}{\Omega_{\rm in}}\right| \left| \frac{v_{g, {\rm in}}}{v_{g, T}} \right|} \, \mathcal{T} \,, \nonumber \\
N &=& - \sqrt{\left| \frac{\Omega_{N}}{\Omega_{\rm in}}\right| \left| \frac{v_{g, {\rm in}}}{v_{g, N}} \right|} \, \mathcal{N} \,.
\label{eq:bare_v_normalized_dispersive}
\end{eqnarray}
The expressions in Eqs.~(\ref{eq:bare_v_normalized}) are simply the restrictions of Eqs.~(\ref{eq:bare_v_normalized_dispersive}) to the non-dispersive case.

It is interesting and useful to note that, in the limit $\omega \to 0$ (where the dispersive and non-dispersive normalizations agree), the normalized amplitude $\mathcal{R}$ and the ratio of normalized amplitudes $\mathcal{N}/\mathcal{T}$ are simply related to the limiting values of $c$ within their respective regions (calculated using methods developed in~\cite{Anderson-Fabbri-Balbinot-2015}):
\begin{equation}
\left|\mathcal{R}\right|^{2} \to \left(\frac{c_{\rm sub}-c_{\rm hor}}{c_{\rm sub}+c_{\rm hor}}\right)^{2} \,, \qquad \left|\frac{\mathcal{N}}{\mathcal{T}}\right|^{2} \to \left(\frac{c_{\rm sup}-c_{\rm hor}}{c_{\rm sup}+c_{\rm hor}}\right)^{2} \,,
\label{eq:low-freq}
\end{equation}
where $c_{\rm sub}$ and $c_{\rm sup}$ are the values of $c$ in the sub- and supercritical regions, respectively, and $c_{\rm hor}$ is its value at the horizon.
Notice that Eqs.~(\ref{eq:low-freq}) are in accordance with the fact that no mode mixing 
occurs if $c$ is constant.


\subsection{Data analysis}

In this section, we provide a 
detailed account of the data analysis used to extract the results presented 
in the Letter. 
This includes technical information and further results complementary to those already presented.

\subsubsection{Background flow}

An original two-dimensional free-surface flow code was developed in order to simulate numerically the transcritical flow studied experimentally in the Letter. 
A projection method is applied to the incompressible variable density Navier-Stokes equations to decouple velocity and pressure unknowns. 
Away from the interfaces (water-air and obstacle-water), partial differential operators (divergence, gradient, Laplacian operator) and nonlinear terms are discretized on a fixed Cartesian grid 
using standard second-order finite difference approximations. 
Several techniques are used to account for the presence of the two interfaces while avoiding the generation of conformal meshes. 
An Immersed Boundary Method~\cite{IBM} enforces the no-slip boundary condition on the rigid obstacle and the free surface evolution is tackled with the Level-Set technique~\cite{LST}.

\begin{figure}[!ht]
\includegraphics[width=\columnwidth]{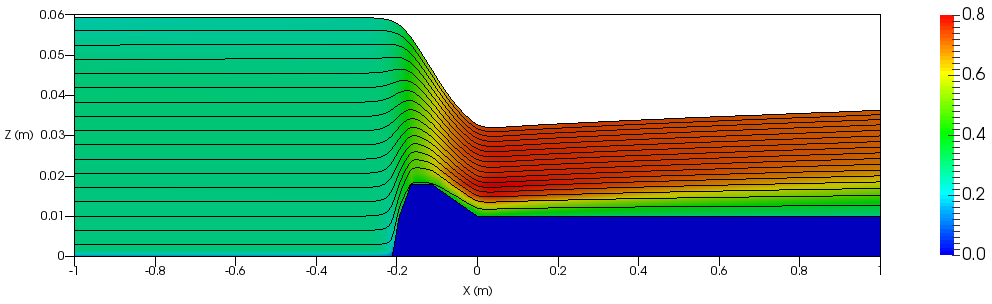} \\
\includegraphics[width=0.95\columnwidth]{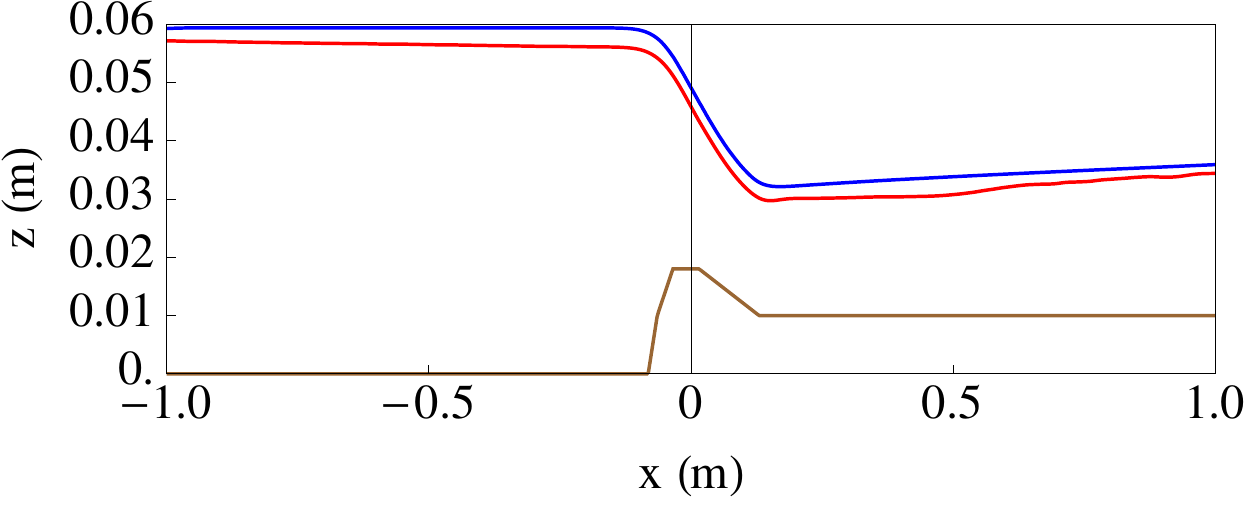}
\caption{Numerical results pertaining to the stationary background flow, which is assumed homogeneous in the transverse direction.  In the upper panel is shown the magnitude of the flow velocity on a color scale (in units of ${\rm m}/{\rm s}$), with its direction lying along the streamlines shown in black.  In the lower panel are shown the numerically calculated free surface (blue) and that found experimentally after averaging $\delta h(x,t)$ in time (red).  The apparent discrepancy averages at $2.60\,{\rm mm}$, which is close to the pixel size used in the simulation ($\Delta z = 2.75 \,{\rm mm}$).
\label{fig:BGnumerical}}
\end{figure}

\begin{figure}
\includegraphics[width=\columnwidth]{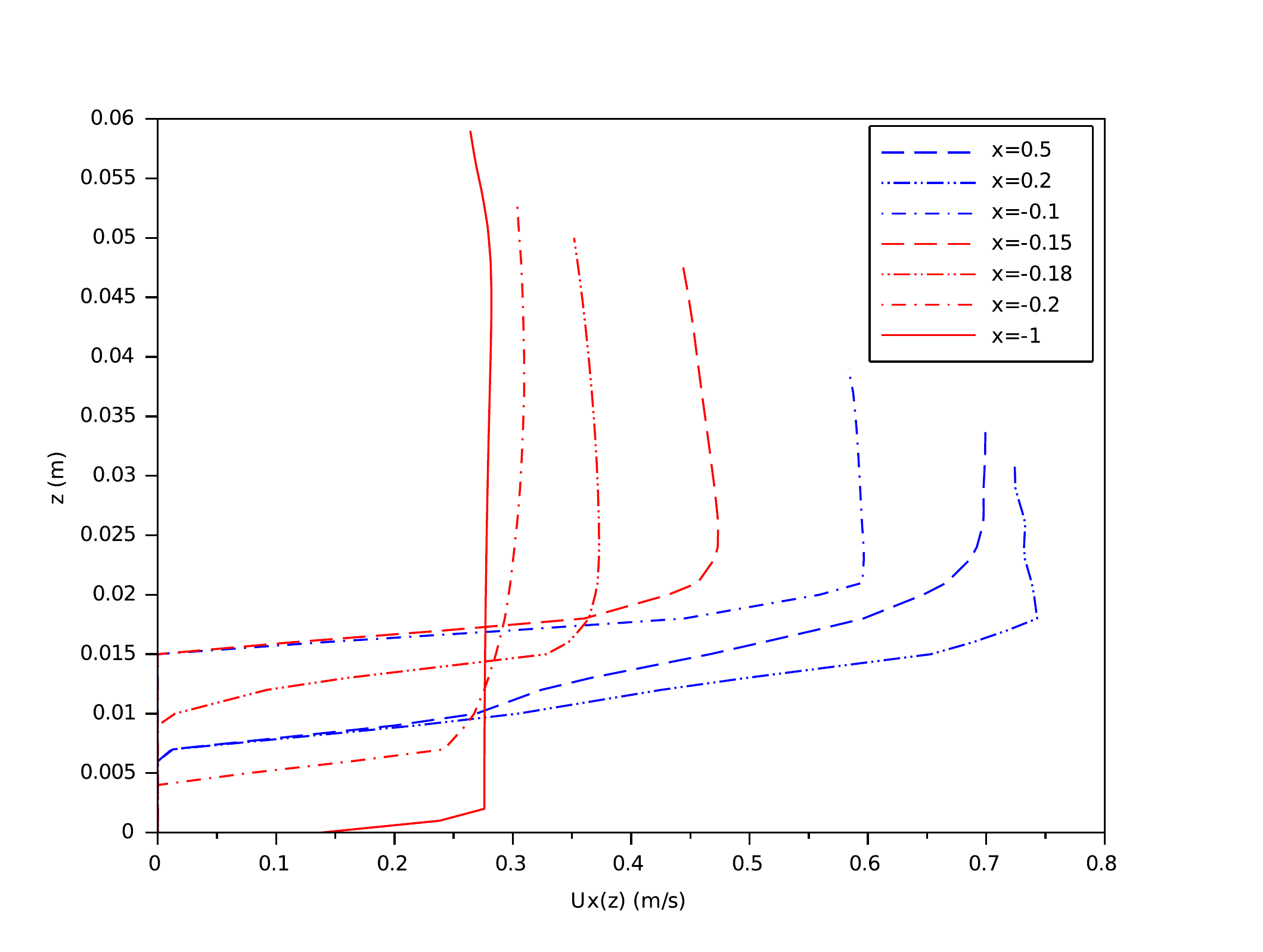}
\caption{The longitudinal component of the flow velocity as a function of the vertical coordinate $z$, at a series of fixed longitudinal coordinates $x$.
\label{fig:BGprofils}}
\end{figure}

In Figures~\ref{fig:BGnumerical} and~\ref{fig:BGprofils} are shown the results of the numerical calculation for the flow across the obstacle, and with a total flow rate close to that used in the experiment.
This assumes that the system is homogeneous in the transverse direction, so that the flow velocity has no dependence on or component along $y$; boundary effects due to the walls of the channel are thus neglected.
In Fig.~\ref{fig:BGnumerical}, the upper panel shows the magnitude of the flow on a color scale (its direction being indicated by the streamlines in black), while the lower panel shows the numerically derived water depth (in blue) on the same graph as the experimental time-averaged water depth (in red).  There is a small apparent discrepancy with a spatial average of $2.60\,{\rm mm}$, which is close to the pixel size used in the simulation ($\Delta z = 2.75\,{\rm mm}$).  Even so, the qualitative similarity between the two can be appreciated.
Figure~\ref{fig:BGprofils} explicitly plots the longitudinal component of the flow velocity as a function of depth at several values of the longitudinal coordinate $x$.

We see that the numerically calculated flow is very close to laminar in the upstream region, before the flow crosses the obstacle.  The standard expressions $v = q/h$ and $c = \sqrt{g h}$ (Eqs.~(3) of the Letter) are thus expected to work very well there.  Moreover, we can expect to be sensitive to small deviations from the idealized theoretical model.  We have already seen that the dispersive corrections to the wave vectors of the reflected waves are visible at high frequencies (see Fig.~3 of the Letter).  We have also noted in the Letter a discrepancy between the measured value of the flow rate $q$, and the fitted value of $q$ assuming the validity of $v = q/h$ in the subcritical region.  The solid curve in Fig.~\ref{fig:BGprofils} indicates that, while $v$ is essentially constant over most of the depth, there is a slight decrease as the free surface is approached.  Since it is the surface value of $v$ that enters the effective metric for surface waves, this behavior near the surface should engender a small decrease in $q_{\rm fit}$ with respect to the measured $q$.  Indeed, in the Letter we find that $q_{\rm fit}$ is smaller than $q$ by about $3 \%$.  We thus believe that the decrease of $v(z)$ near the surface is responsible for the discrepancy between $q$ and $q_{\rm fit}$.

The numerical results also indicate that we should expect somewhat larger discrepancies from the standard predictions for 
$v$ and $c$ once we enter the supercritical (downstream) region, for there the variation of the flow velocity with depth is more significant.  This is again what is observed: the extracted value of $v$ is considerably larger than $q/h$, which is consistent with the appearance of a significant boundary layer at the bottom of the flow where $v$ is smaller than it is near the surface.

\begin{figure}
\includegraphics[width=0.9\columnwidth]{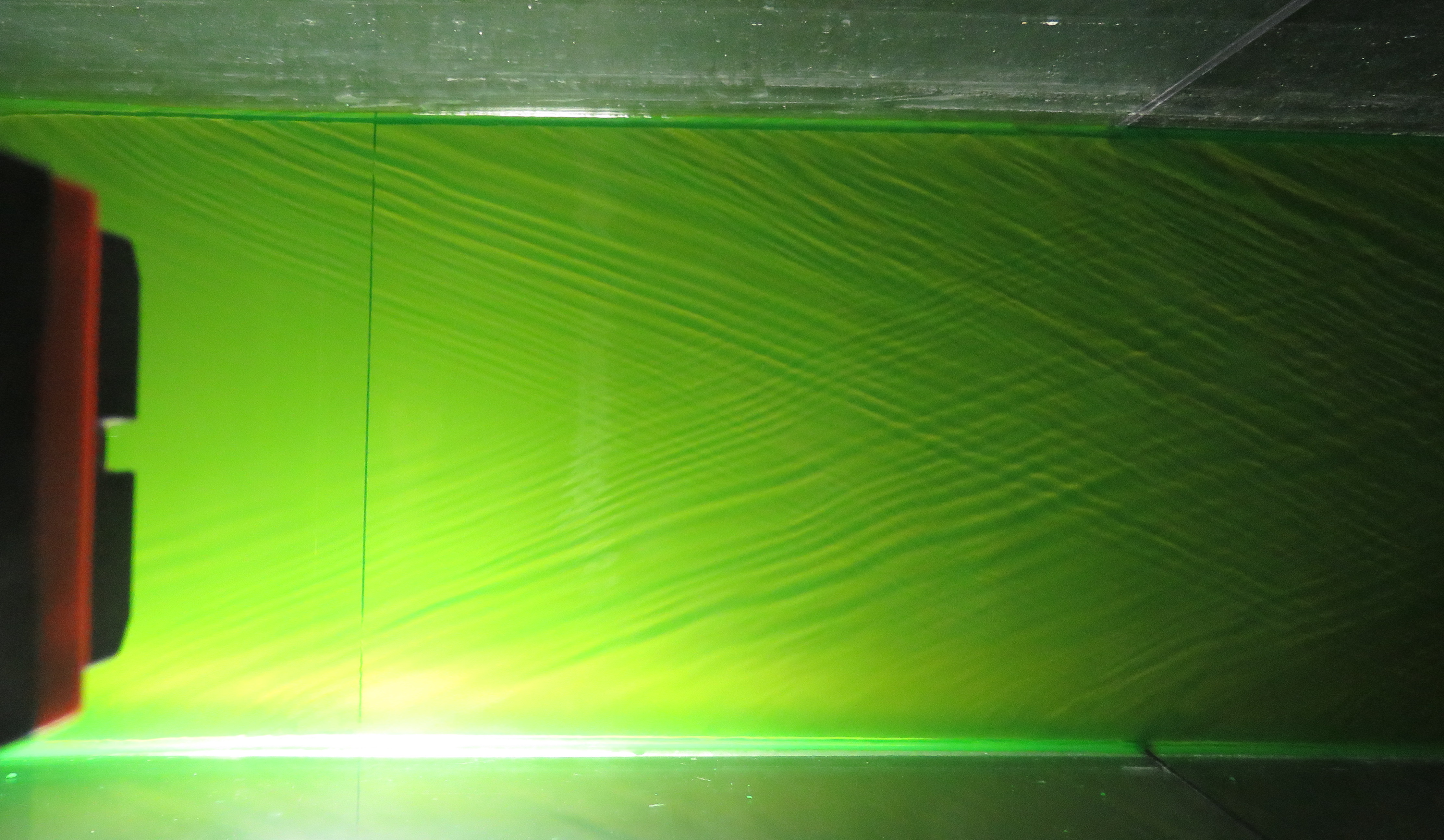}
\caption{Aerial view of the water surface in the supercritical region.  The side wakes -- emanations from the walls of the channel -- are clearly visible.
\label{fig:SideWakes}}
\end{figure}

It is important to emphasize the two-dimensional nature of the numerical calculations used to produce Fig.~\ref{fig:BGnumerical}.  That is, the flow velocity is assumed to be completely independent of the transverse direction.  This means that any boundary effects due to the walls of the flume have not been taken into account.  One such effect, visible to the naked eye, is shown in the photo of Fig.~\ref{fig:SideWakes}: a quasi-stationary wake pattern emanating from the walls of the flume in the supercritical region.  By ``quasi-stationary'' we mean that it is not strictly time-dependent, with positions of the crests and troughs showing a considerable amount of jitter that is noticeable by eye and which will contribute to the free-surface fluctuations $\delta h(t,x)$.  This is likely the source of the increased noise level in the supercritical region. 

\subsubsection{Extracting wave vectors and amplitudes from the experimental data}

\begin{figure}
\includegraphics[width=0.95\columnwidth]{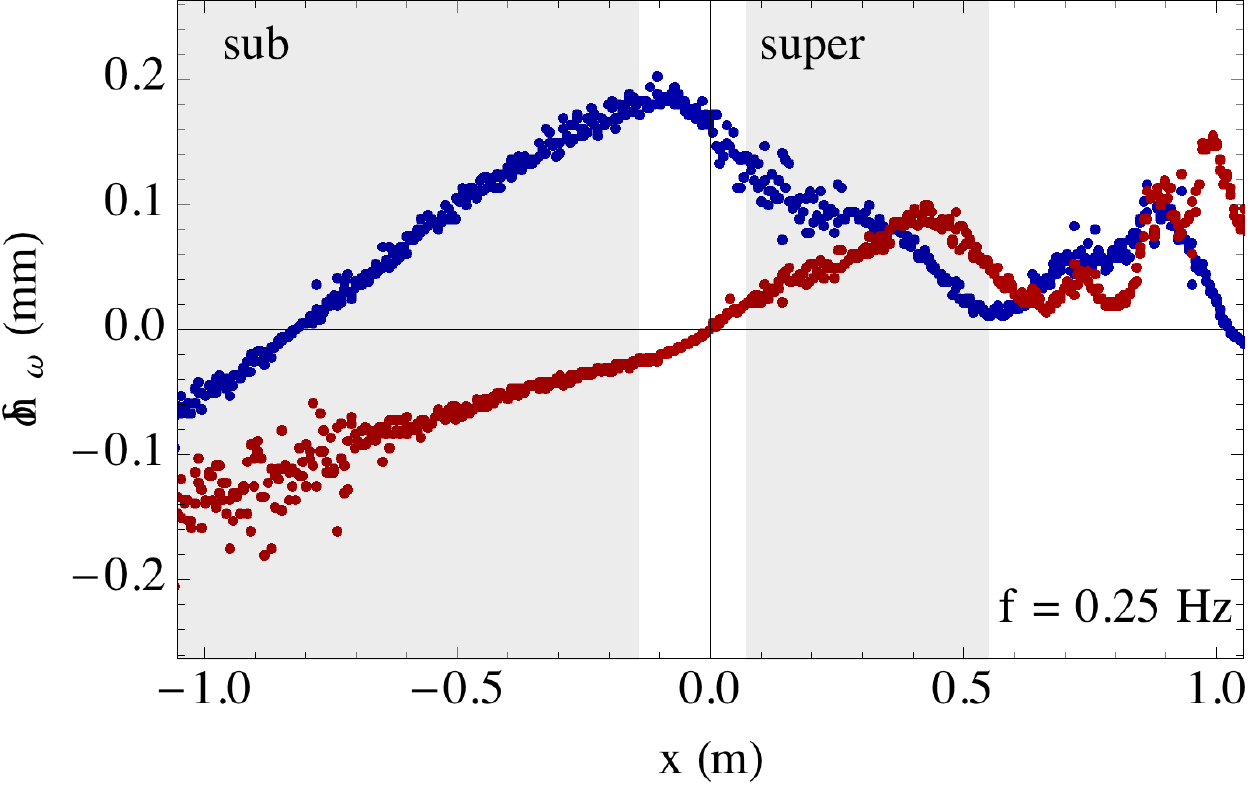} \\ \includegraphics[width=0.95\columnwidth]{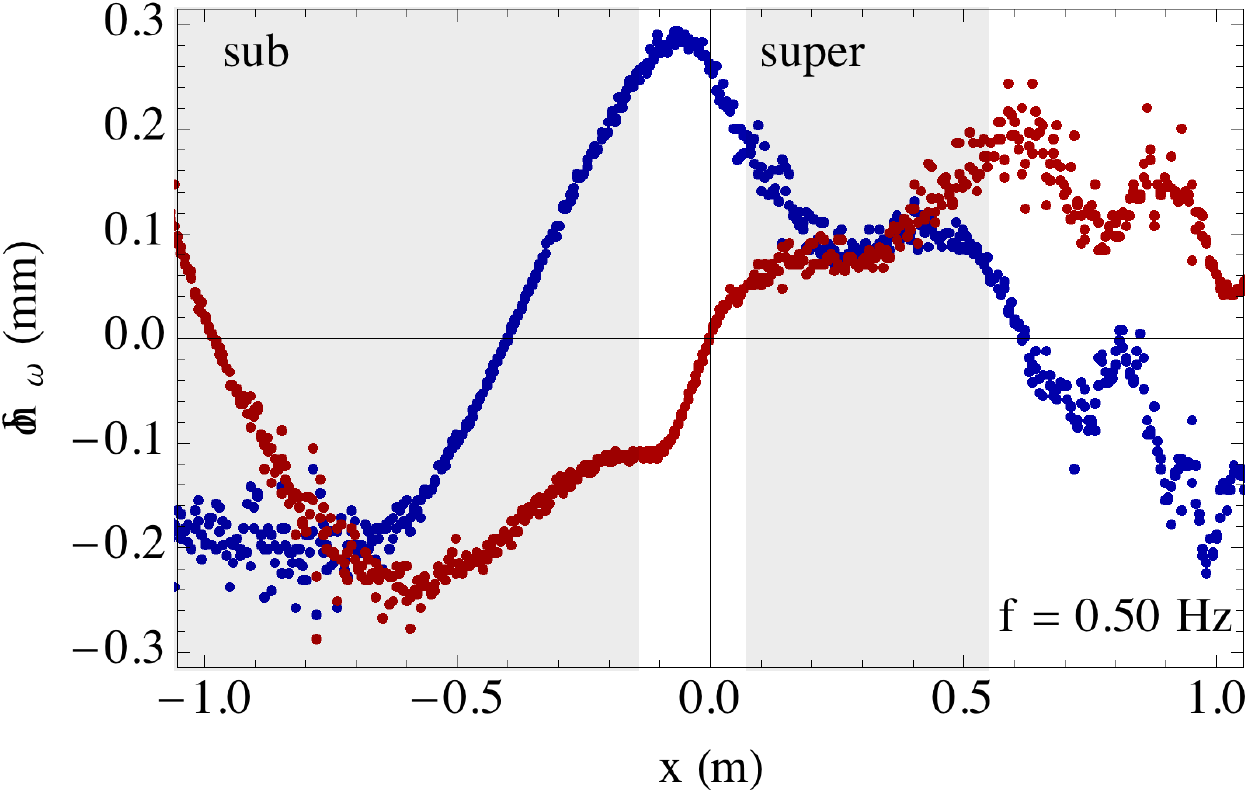}
\caption{Extracted waveform $\delta h_{\omega}(x)$, found by Fourier transforming the data in time at the stimulation frequency of the wave maker.  The chosen runs are at frequencies $0.25\,{\rm Hz}$ (upper panel) and $0.5\,{\rm Hz}$ (lower panel), with a wave maker amplitude $A_{\rm wm} = 0.5\,{\rm mm}$.  Blue and red data points correspond to the real and imaginary parts, respectively, of $\delta h_{\omega}(x)$. 
The sub- and supercritical regions, where fitting to a sum of two plane waves is performed, are shaded.
\label{fig:delta-h}}
\end{figure}

Measurements are made of the free-surface deformation, $\delta h (t,x)$.
We suppose that in each run this 
can be written as
\begin{linenomath}\begin{eqnarray}
\!\!\!\!\!\!\!\! \delta h(t,x) & = & {\rm Re}\left\{ \delta h_{\omega}(x) \, e^{-i\omega t}\right\} + {\rm noise} \nonumber \\
& = & \frac{1}{2} \, \delta h_{\omega}(x) \, e^{-i\omega t} + \frac{1}{2} \, \delta h^{\star}_{\omega}(x) \, e^{i \omega t} + {\rm noise} \,,
\label{eq:delta-h_complex}
\end{eqnarray}\end{linenomath}
where $\omega$ is the frequency of stimulation by the wave maker.
$\delta h_{\omega}(x)$ can thus be extracted by taking the Fourier component at 
frequency 
$\omega$:
\begin{linenomath}\begin{equation}
\delta h_{\omega}(x) \approx \frac{2}{N} \sum_{j=1}^{N} e^{i \omega t_{j}} \delta h\left(t_{j}\,,\,x\right) \,.
\end{equation}\end{linenomath}
Here, the $t_{j}$ are the discrete times at which the data has been recorded.  The total number of such times is $N_{t}=8192$, for a total duration of $800\,{\rm s}$.  
Examples of the extracted waveform, at two different frequencies, are shown in Fig.~\ref{fig:delta-h}.

\begin{figure*}
\includegraphics[width=0.4\textwidth]{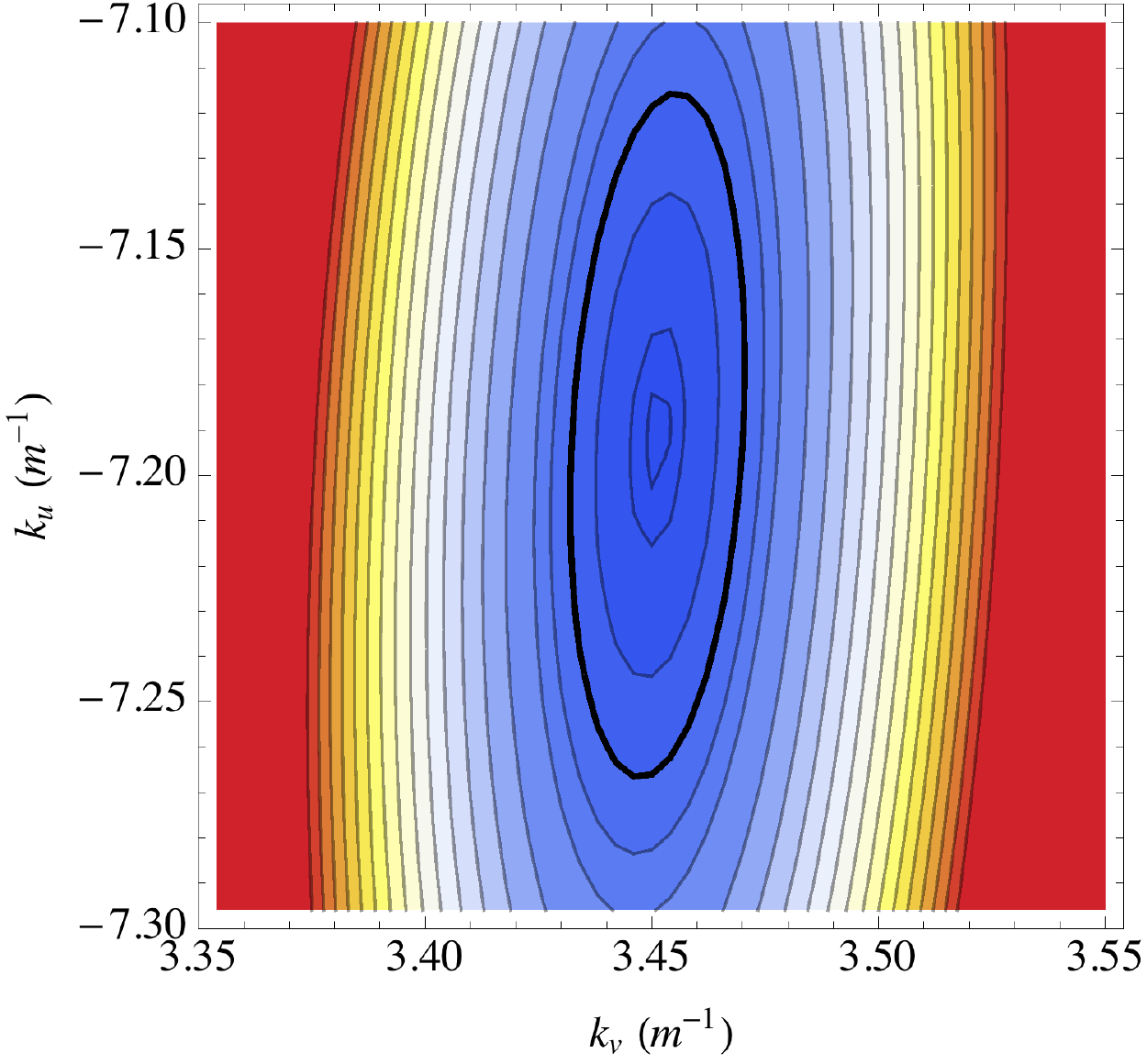} \, \includegraphics[width=0.4\textwidth]{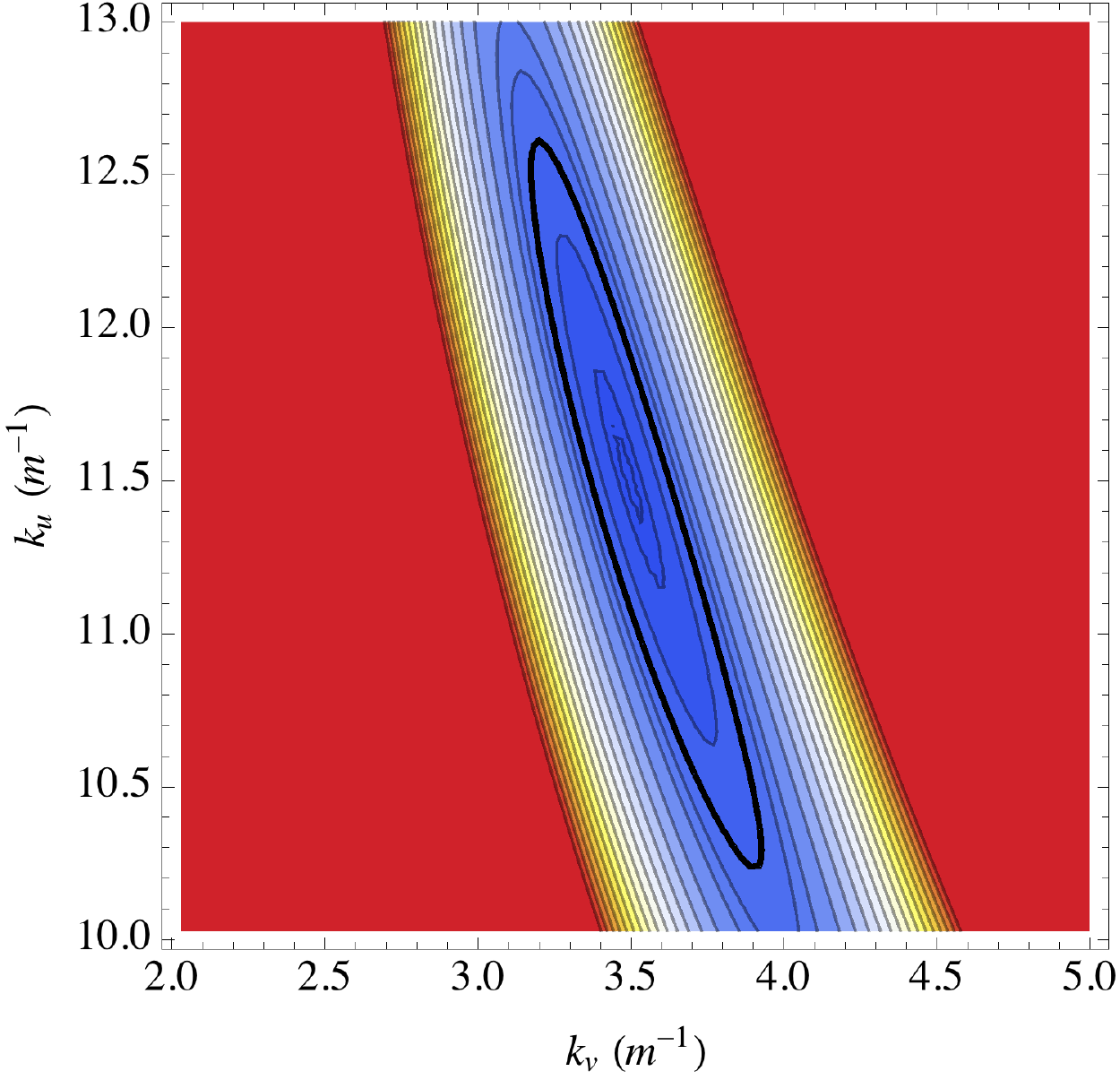} \, \includegraphics[width=0.0533\textwidth]{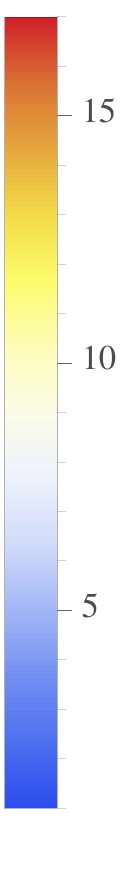}
\caption{Reduced $\chi^{2}$ in the 2D wave vector plane, shown on a color scale.  The values are shown for the run with $A_{\rm wm} = 0.25\,{\rm mm}$ and $f = 0.55\,{\rm Hz}$, in the subcritical region (left panel) and in the supercritical region (right panel).  For each wave vector pair, $\left(k_{v},\,k_{u}\right)$, a linear fit of the wave amplitudes is performed and the reduced $\chi^{2}$ is extracted.  The point where this reaches a minimum value (defined to be $1$) represents the best fit, while the locus where it increases to $2$ (indicated by the thick black contour) defines the standard error. 
The projections of this region onto each of the axes give the standard errors in $k_{v}$ and $k_{u}$ shown in Fig.~3 of the Letter.
\label{fig:ellipse_A025_f055}}
\end{figure*}

With the complex waveform $\delta h_{\omega}(x)$ having been extracted, we focus our attention on the near-homogeneous regions on either side of the horizon.  There, the dispersion relation $\omega(k)$ is essentially independent of position and there should be almost no scattering between the various waves.  Therefore, the complex waveform should be expressible as a sum of plane waves:
\begin{linenomath}\begin{equation}
\delta h_{\omega,{\rm hom}}(x) = A_{1} \, e^{i k_{1} x} + A_{2} \, e^{i k_{2} x} \,.
\label{eq:plane_wave_decomp}
\end{equation}\end{linenomath}
Since we wish to fit the wave vectors as well as the amplitudes, the fit 
is performed using the Mathematica function ``NonlinearModelFit'', which also provides estimates for the errors in the fitting parameters.  
The procedure used both for the fitting and the error estimation can be illustrated as in Fig.~\ref{fig:ellipse_A025_f055}.
We consider a section of the $\left(k_{1},\,k_{2}\right)$-plane, and for each point on this plane we may perform a linear fit of the amplitudes $A_{1}$ and $A_{2}$ at the given values of $k_{1}$ and $k_{2}$.
This fit is precisely defined via the minimization of $\chi^{2}$ (the sum of the squared residuals): 
\begin{equation}
\chi^{2} = \sum_{j=1}^{N_{x}} \frac{\left| \delta h_{\omega}\left(x_{j}\right) - \delta h_{\omega,{\rm hom}}\left(x_{j}\right) \right|^{2}}{\sigma^{2}} \,,
\end{equation}
where we have assumed that the error $\sigma$ is independent of position. 
For each $\left(k_{1},\,k_{2}\right)$, minimizing $\chi^{2}$ returns the best fit values of $A_{1}$ and $A_{2}$.
The error $\sigma$ is estimated by assuming the {\it reduced} $\chi^{2}$, i.e., $\chi^{2}$ per degree of freedom (here $\chi^{2}/\left(2 N_{x} - 6\right)$, an extra factor of $2$ appearing since the function is complex-valued), is equal to $1$ at the minimum.
Shown in Fig.~\ref{fig:ellipse_A025_f055} are contour plots of the reduced $\chi^{2}$, and we clearly see a well-defined local minimum at some point $\left(k_{1},\,k_{2}\right)$.
The errors on $k_{1}$ and $k_{2}$ are estimated by considering the contour where the reduced $\chi^{2}$ is equal to $2$ (shown in thick black in Fig.~\ref{fig:ellipse_A025_f055}). 
As for any normal probability distribution, it is estimated that the actual $\left(k_{1},\,k_{2}\right)$ lies in the region bounded by this contour with a probability of about $68\%$, and the errors on $k_{1}$ and $k_{2}$ 
are just the projected widths 
of this region on the $k_{1}$- and $k_{2}$-axes.

Note that the errors on $k_{1}$ and $k_{2}$ defined by this ellipse will feed into the errors associated with the amplitudes; for, while the errors on the amplitudes for each {\it fixed} point $\left(k_{1},\,k_{2}\right)$ are defined by the standard linear fitting procedure and may not be especially large, the corresponding best fit values of the amplitudes may vary quite a lot over the length of the ellipse in the $\left(k_{1},\,k_{2}\right)$-plane.  This variation must then be included in the error on the amplitudes, and may turn out to be the dominant contribution.  This appears to be what happens when the errors on $T$ and $N$ become large around $f = 0.5\,{\rm Hz}$ in Fig.~4 of the Letter: the region of the $\left(k_{1},\,k_{2}\right)$-plane where the wave vectors are likely to lie turns out to be rather long (see the right panel of Fig.~\ref{fig:ellipse_A025_f055}), and the variation of $T$ and $N$ along it is rather large.

It is also to be noted that some trial and error is required when performing a nonlinear fit of the type described here, with the wave vectors being included as fitting parameters.
This is because (unlike in a linear fit) a local minimum of $\chi^{2}$ is not uniquely defined, and is therefore not guaranteed to be a global minimum.
We therefore look in the vicinity of the expected value of $\left(k_{v},\,k_{u}\right)$, and the minimum closest to it selected.
That said, when working in the supercritical region, there are a few cases where a larger-than-expected value of $k_{u}$ appears to be unavoidable; these are the outliers circled in Fig.~3 of the Letter.

\subsubsection{Comparing theoretical and measured scattering amplitudes}

\begin{figure}
\includegraphics[width=0.8\columnwidth]{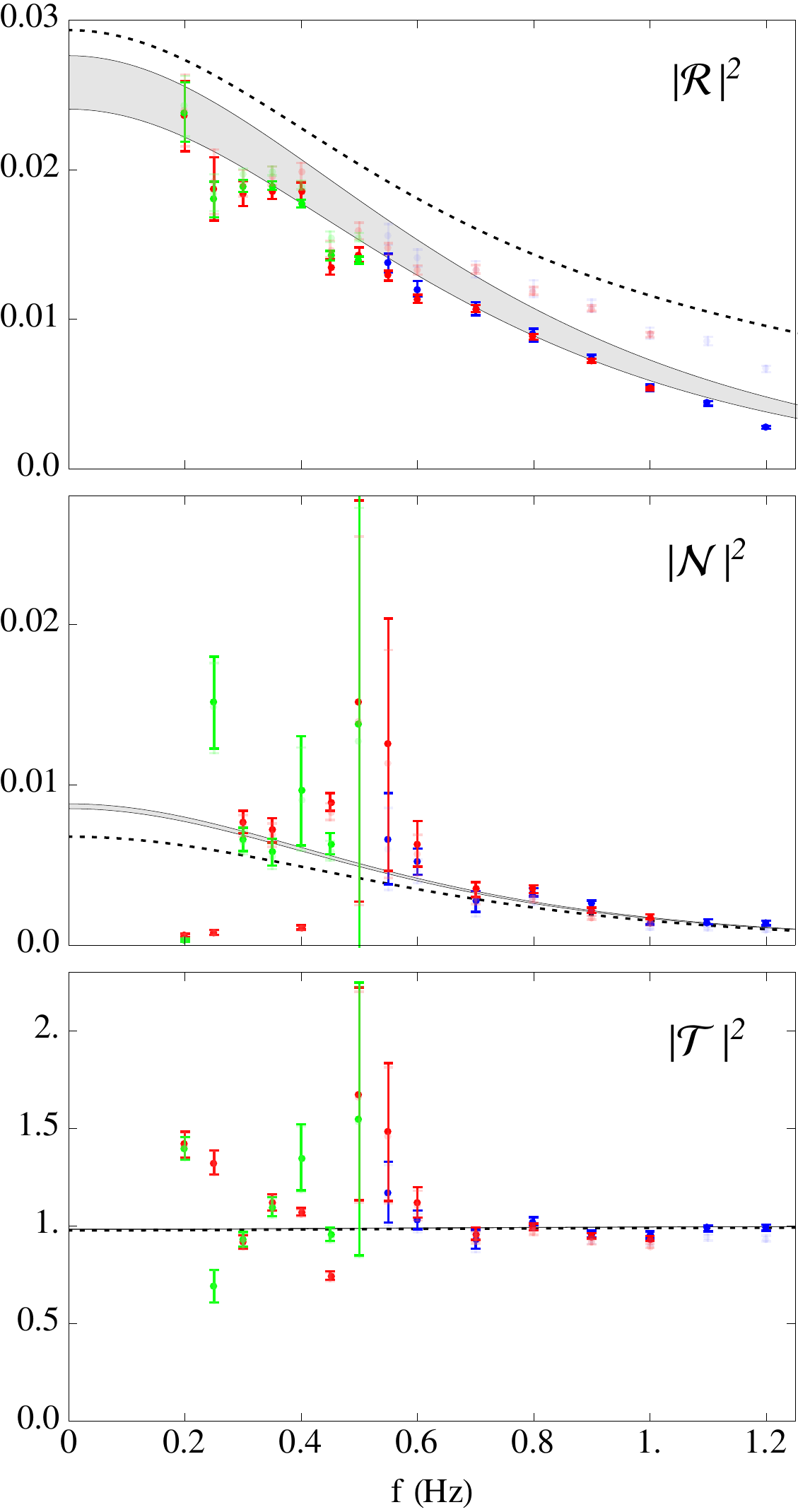}
\caption{Normalized scattering amplitudes.  The differently colored data points correspond to the different amplitudes of the wave maker: $A_{\rm wm} = 0.25\,{\rm mm}$ (blue), $0.5\,{\rm mm}$ (red) and $1\,{\rm mm}$ (green).  Normal shading of data points indicates the use of dispersive normalization factors, while light shading is for the results obtained when normalizing according to the non-dispersive theory.  The dotted curves give the predictions of the hydrodynamical theory when using the standard expressions for $v$ and $c$ (Eqs.~(3) of the Letter), 
while the gray shaded regions correspond to the predictions of the refined model using Eqs.~(\ref{eq:bernoulli}).
\label{fig:normalized}}
\end{figure}

In Fig.~4 of the Letter, we have shown the ``bare'' (unnormalized) scattering amplitudes, extracted directly from the data since they are simply ratios of the measured outgoing wave amplitudes to that of the incident wave.  In that case, the theoretical predictions yielded by Eq.~(\ref{eq:wave_eqn_amplitudes}) are converted into ``bare'' amplitudes using Eqs.~(\ref{eq:bare_v_normalized_dispersive}); the conversion factors, which take dispersive corrections into account, are evaluated for the values of $v$ and $c$ extracted in Fig.~3 of the Letter.  Here instead, we adopt the inverse approach (see Fig.~\ref{fig:normalized}): the normalized amplitudes directly calculated using Eq.~(\ref{eq:wave_eqn_amplitudes}) are shown, and the ``bare'' amplitudes extracted from the data are converted into normalized amplitudes according to the inverse forms of Eqs.~(\ref{eq:bare_v_normalized_dispersive}).  The conversion factors are once again dispersive and evaluated for the extracted $v$ and $c$, but for comparison we also show (with fainter data points) the results we would obtain using the non-dispersive conversion formulae of Eqs.~(\ref{eq:bare_v_normalized}).  
For clarity, the three scattering amplitudes are shown on separate panels, with the colors of the data points indicating the corresponding amplitude of the wave maker oscillations: $A_{\rm wm} = 0.25\,{\rm mm}$ (blue), $0.5\,{\rm mm}$ (red) and $1\,{\rm mm}$ (green).  In theory these should satisfy the unitarity relation~(\ref{eq:unitarity}), but since $\left|\mathcal{R}\right|^{2}$ and $\left|\mathcal{N}\right|^{2}$ are a lot smaller than $\left|\mathcal{T}\right|^{2}$, the sum on the left-hand side of~(\ref{eq:unitarity}) is indistinguishable from $\left|\mathcal{T}\right|^{2}$; we are thus unable to demonstrate that $\left|\mathcal{R}\right|^{2}$ and $\left|\mathcal{N}\right|^{2}$ contribute to the unitarity relation in the expected way. 

\begin{figure}
\includegraphics[width=0.95\columnwidth]{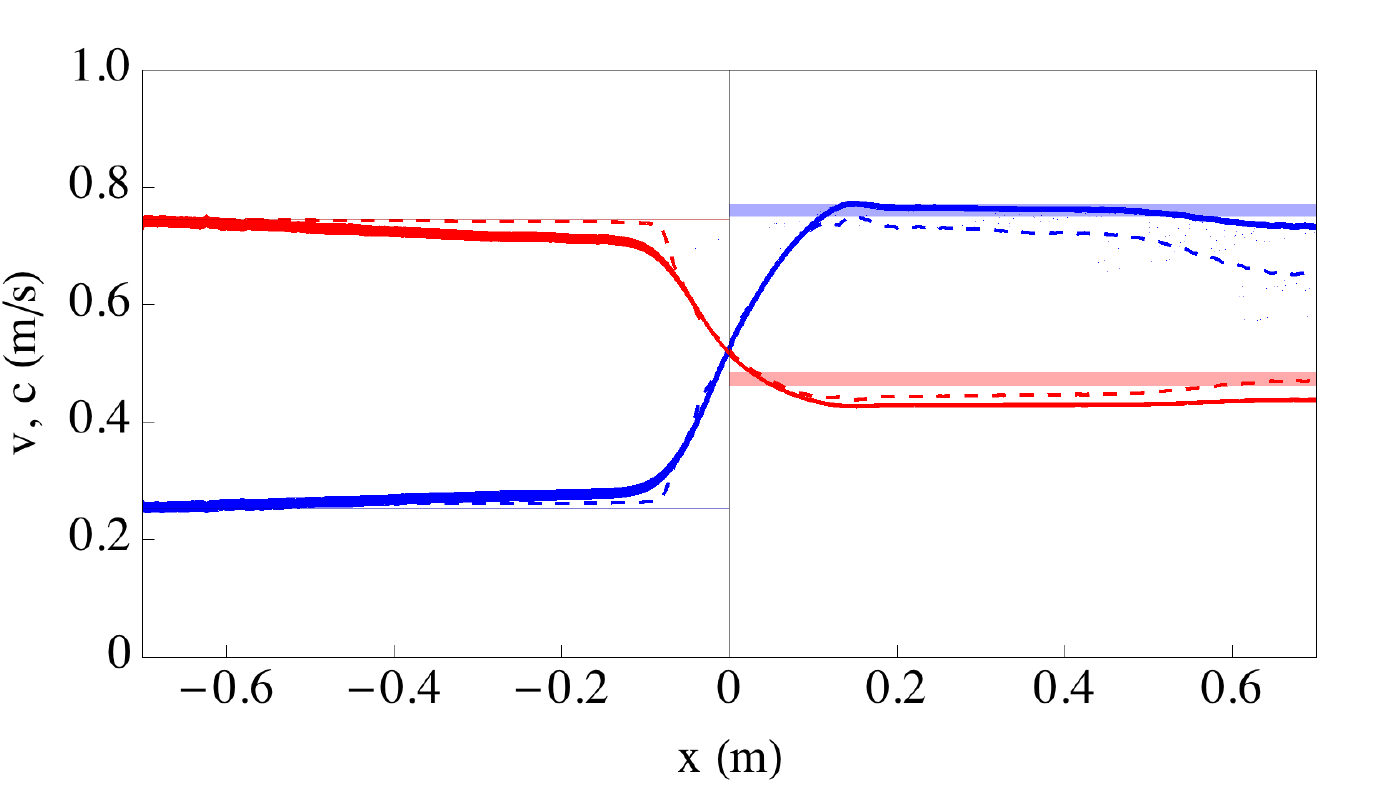}
\caption{Profiles of $v$ (blue) and $c$ (red), according to Eqs.~(3) of the Letter (dashed) and to Eqs.~(\ref{eq:bernoulli}) of this Supplemental Material (solid).  The shaded regions indicate the values of $v$ and $c$ extracted in the near-homogeneous regions (see Fig.~3 of the Letter).
\label{fig:flow_standard_v_bernoulli}}
\end{figure}

For each scattering amplitude, two theoretical predictions are shown which differ in their description of the background.
The dotted curves in Fig.~\ref{fig:normalized} show the predictions 
when using the standard expressions for $v(x)$ and $c(x)$ given by Eqs.~(3) of the Letter.  
Although these are not far from the observed values, there is a particularly noticeable discrepancy in the $\left|\mathcal{R}\right|^{2}$ coefficient.
To try and improve the comparison between theory and experiment,
we also consider the following expressions~\cite{Coutant-Parentani-2014}~\footnote{Starting from Eqs.~(3) of~\cite{Coutant-Parentani-2014}, Eqs.~(\ref{eq:bernoulli}) are derived by assuming quasi-flatness of the surface so that $v_{x} \approx v$, and by neglecting $\partial_{y}v$ near the surface so that $g_{\rm eff} \approx g$.}: 
\begin{eqnarray}
\frac{1}{2} v^{2}(x) + g z_{\rm surface}(x) &=& \frac{1}{2} v_{\rm as}^{2} + g h_{\rm as} \,, \nonumber \\ 
c^{2}(x) &=& g h_{\rm as} \frac{v_{\rm as}}{v(x)} \,.
\label{eq:bernoulli}
\end{eqnarray}
The corresponding $v(x)$ and $c(x)$ are shown in Fig.~\ref{fig:flow_standard_v_bernoulli}, along with the standard profiles of Eqs.~(3) of the Letter and the extracted values of $v$ and $c$ in the near-homogeneous regions. 
The asymptotic values $h_{\rm as}$ and $v_{\rm as}$ entering Eqs.~(\ref{eq:bernoulli}) are taken to be those extracted in the subsonic region: $h_{\rm sub} = 56.7 \pm 0.1 \,{\rm mm}$ and $v_{\rm sub} = 25.3 \pm 0.1 \,{\rm cm}/{\rm s}$, with the errors on $h_{\rm sub}$ and $v_{\rm sub}$ allowing us to define a prediction ``band'' rather than a single well-defined curve.
Interestingly, the profile of $v(x)$ given by Eqs.~(\ref{eq:bernoulli}) is (unlike that of Eqs.~(3) of the Letter) wholly consistent with the value of $v_{\rm sup}$ extracted in the supersonic region. 
By contrast, the profile of $c(x)$ is slightly further from the extracted value of $c_{\rm sup}$ than that given by Eqs.~(3) of the Letter.
It should be noted that $v(x)$ and $c(x)$ of Eqs.~(\ref{eq:bernoulli}) are more sensitive to small changes of $z_{\rm surface}(x)$ in the subsonic region, and show a bit more variation there than Eqs.~(3) of the Letter (a noticeable slope can be seen in Fig.~\ref{fig:flow_standard_v_bernoulli}).  Therefore, when using Eq.~(\ref{eq:wave_eqn_amplitudes}) to produce theoretical scattering amplitudes, we stop the integration just outside the inhomogeneity engendered by the obstacle (specifically at $x = \pm 0.2 \, {\rm m}$)~\footnote{The residual variation of $v$ and $c$ when moving away from the obstacle generates some oscillations in the predictions for the scattering amplitudes, qualitatively similar to those seen at low frequencies in $\left|\mathcal{R}\right|^{2}$ (see the top panel of Fig.~\ref{fig:normalized}).  However, we were unable to get quantitative agreement concerning the period of the observed oscillations.  Given the sensitivity of the prediction for $\left|\mathcal{R}\right|^{2}$ to small variations in $h_{\rm sub}$ and $v_{\rm sub}$, it may be that a greater precision in the measurement of the background is required to get agreement with the details of the observed $\left| \mathcal{R} \right|^{2}$.}.

Returning to Fig.~\ref{fig:normalized}, it is clear that the combination of using dispersive normalization factors to convert the ``bare'' amplitudes to normalized ones, and the use of Eqs.~(\ref{eq:bernoulli}) for $v(x)$ and $c(x)$ rather than Eqs.~(3) of the Letter, yields the most favorable comparison between theory and experiment.
The respective differences are most noticeable in $\left|\mathcal{R}\right|^{2}$, which shows the greatest sensitivity both to dispersive normalization factors (as might have been expected) and to the precise forms of $v(x)$ and $c(x)$.
The other amplitudes also show some improvement: $\left|\mathcal{N}\right|^{2}$ is rather insensitive to dispersion but at low frequencies is more compatible with the refined versions of $v(x)$ and $c(x)$, while $\left|\mathcal{T}\right|^{2}$ is rather insensitive to changes of the background but shows some improvement at high frequencies when dispersion is included in the normalization factors.
Note that dispersion plays a minimal role here, entering only in the normalization factors but being absent from the predictions of the normalized scattering amplitudes.  It thus seems to play little role in the scattering, at least until we reach the highest frequencies probed and we start to see the measured $\left|\mathcal{R}\right|^{2}$ falling further away from the theoretical prediction.

\subsubsection{Background noise and harmonics}

The fitting performed in deriving the results above 
assumes that the only element of $\delta h(t,x)$ is the linear perturbation oscillating at the stimulated frequency $\omega$, i.e. that $\delta h(t,x) = \mathrm{Re}\left\{ \delta h_{\omega}(x) e^{-i \omega t} \right\}$.
In fact, there are other contributions, and it is useful to be aware of them and their relative importance compared with the main signal. 

\begin{figure*}
\includegraphics[width=0.35\textwidth]{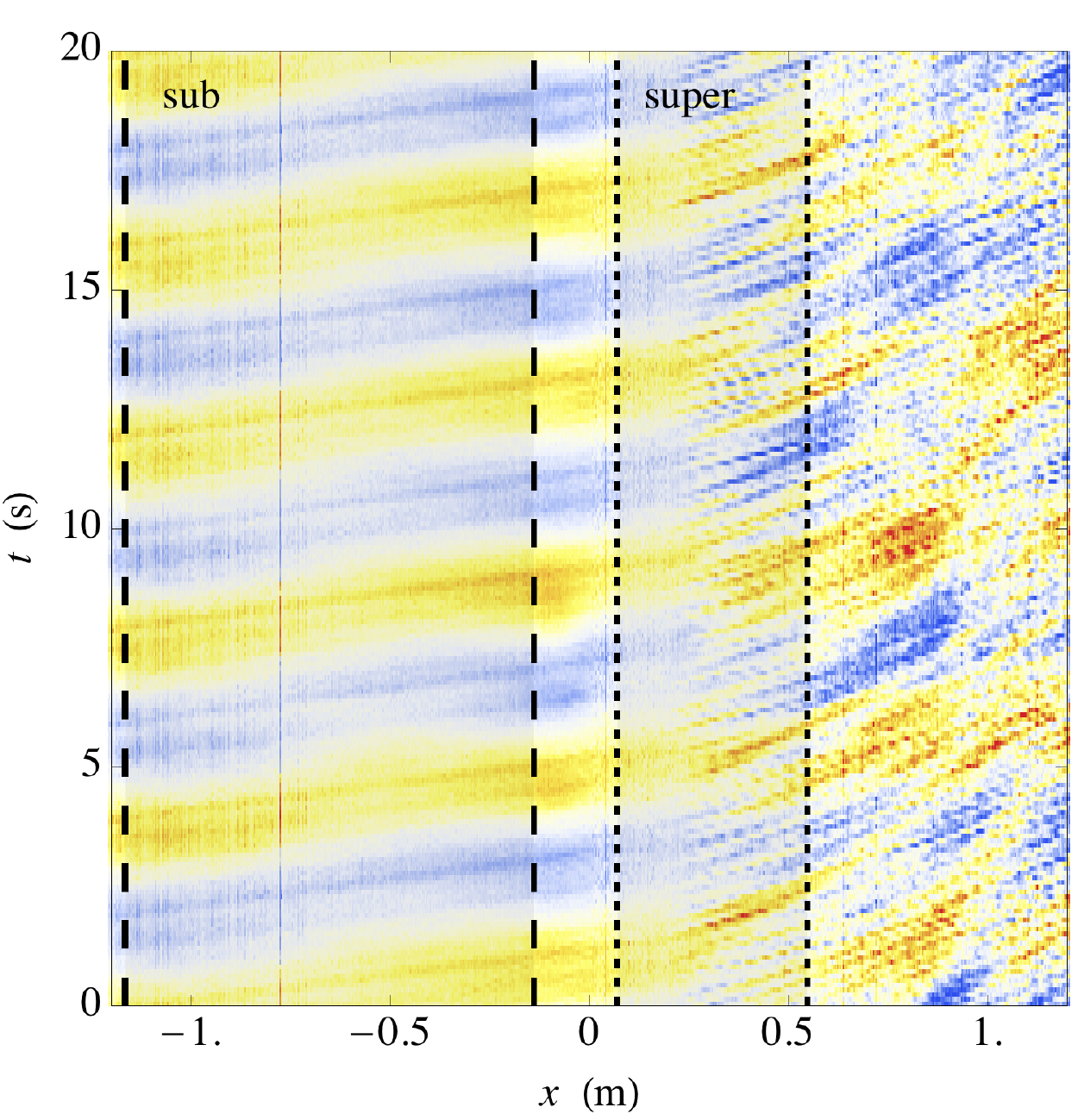} \, \includegraphics[width=0.35\textwidth]{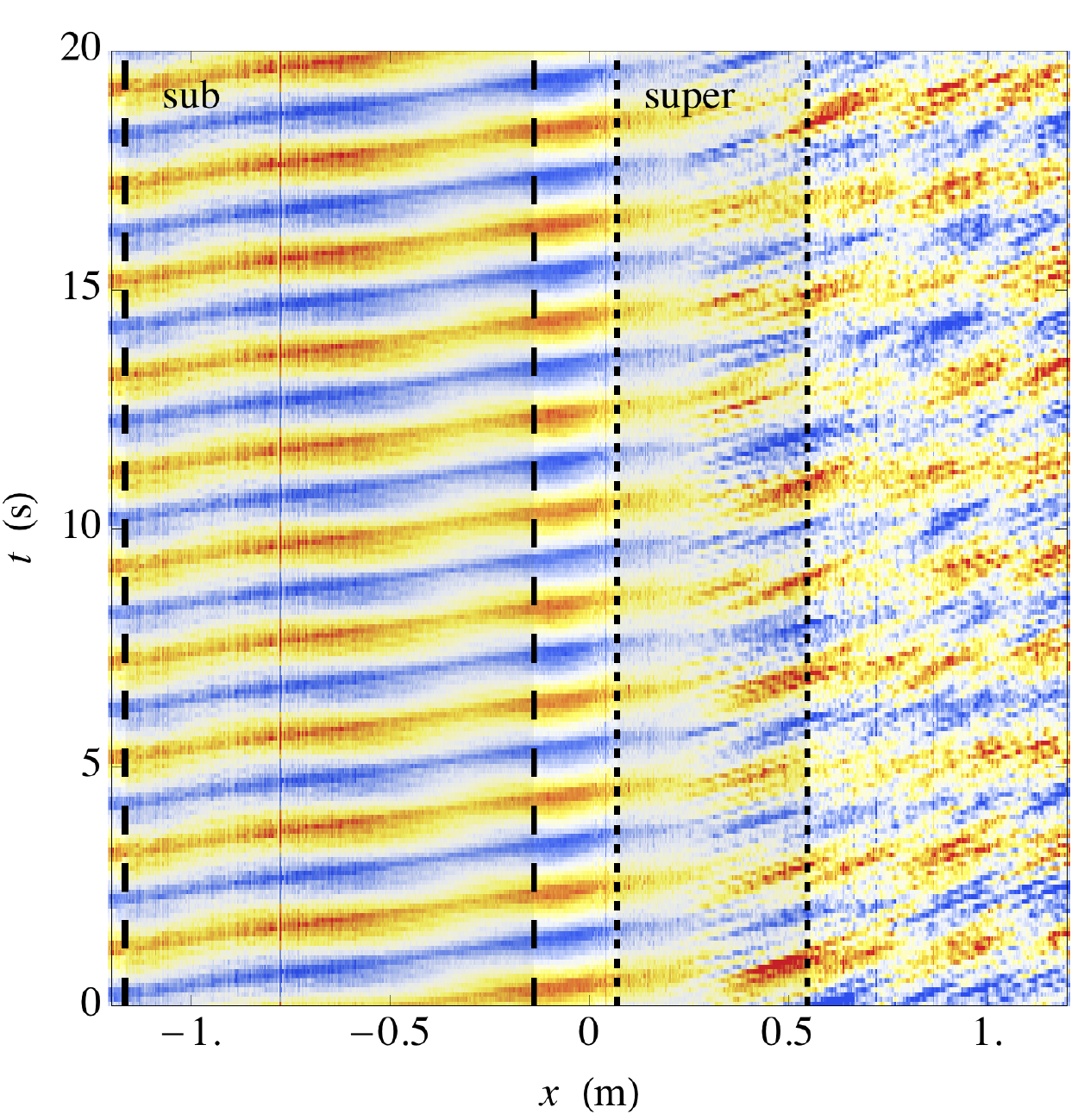} \, \includegraphics[width=0.0544\textwidth]{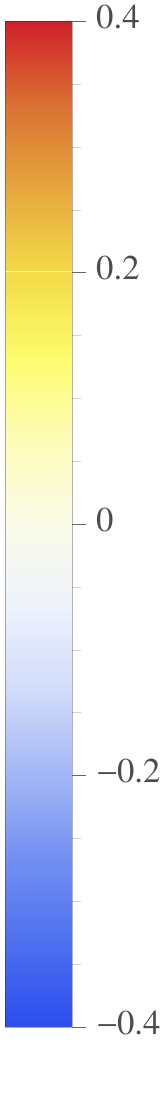}
\caption{Space-time diagrams showing $\delta h(x,t)$ for two different experimental runs, with the wave maker frequency equal to $0.25\,{\rm Hz}$ (left panel) and $0.5\,{\rm Hz}$ (right panel), and the wave maker amplitude fixed at $0.5\,{\rm mm}$.  $\delta h$ is plotted on a color scale as indicated by the color bar on the right (the values are in $\mathrm{mm}$).  The black 
vertical lines correspond to the limits of the near-homogeneous regions on the subcritical (dashed) 
and supercritical (dotted) 
sides, as also illustrated in Fig.~1 of the Letter and in Fig.~\ref{fig:delta-h} of this Supplementary Material. 
On the subcritical side, the data is clean and regular, and in the right plot we can see variations of the amplitude of the oscillations due to interference between the incident and reflected waves.  On the supercritical side, however, the data clearly becomes much noisier, and this growth appears to occur precisely in the near-homogeneous region delineated by the dotted 
vertical lines.
\label{fig:STdiagram_A050}}
\end{figure*}

In Figure~\ref{fig:STdiagram_A050} are shown space-time diagrams plotting $\delta h(t,x)$, for frequencies $f = 0.25\,{\rm Hz}$ (left panel) and $f = 0.5\,{\rm Hz}$ (right panel) and a wave maker amplitude of $0.5\,{\rm mm}$. 
The sub- and supercritical near-homogeneous regions are delineated by the dashed and dotted 
vertical lines, respectively. Of particular notice is the noisy character of the data in the supercritical region, with this noise appearing to begin precisely in the near-homogeneous region where the fitting is performed.  The data in the subcritical region is, by contrast, very clean and regular.

\begin{figure*}
\includegraphics[width=0.35\textwidth]{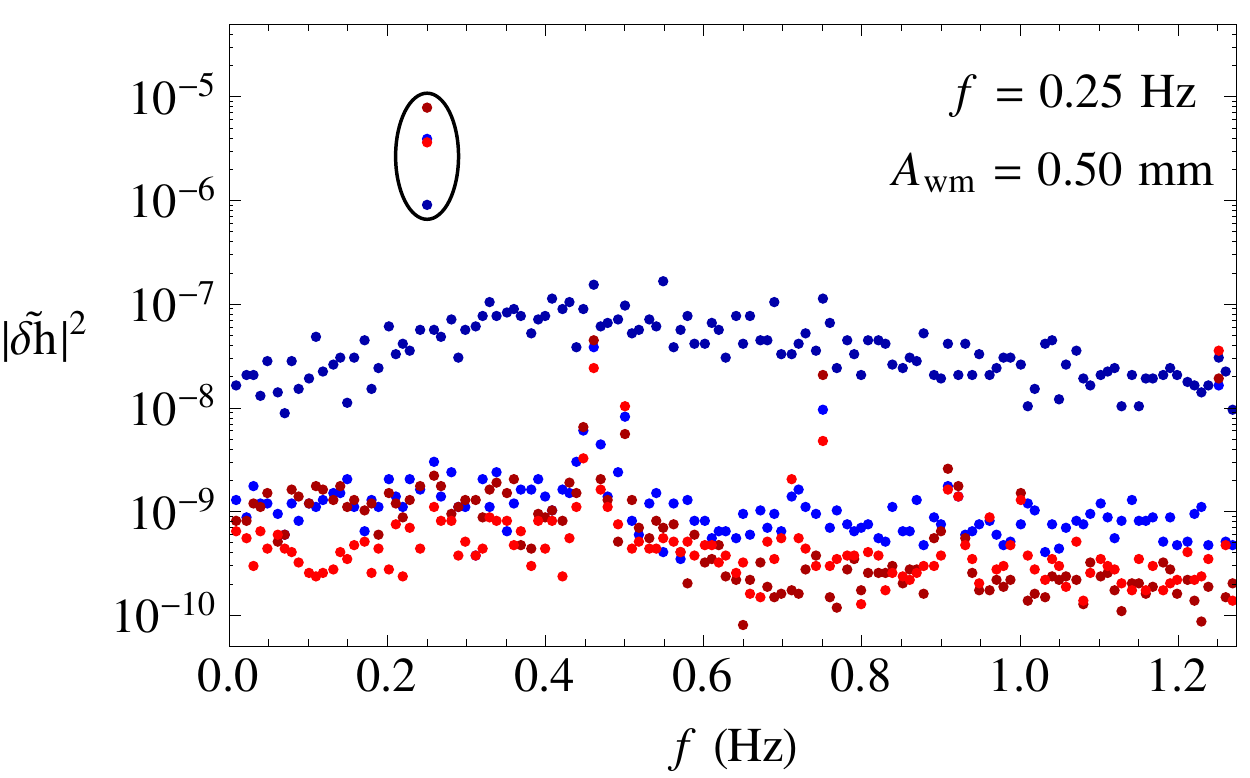} \, \includegraphics[width=0.35\textwidth]{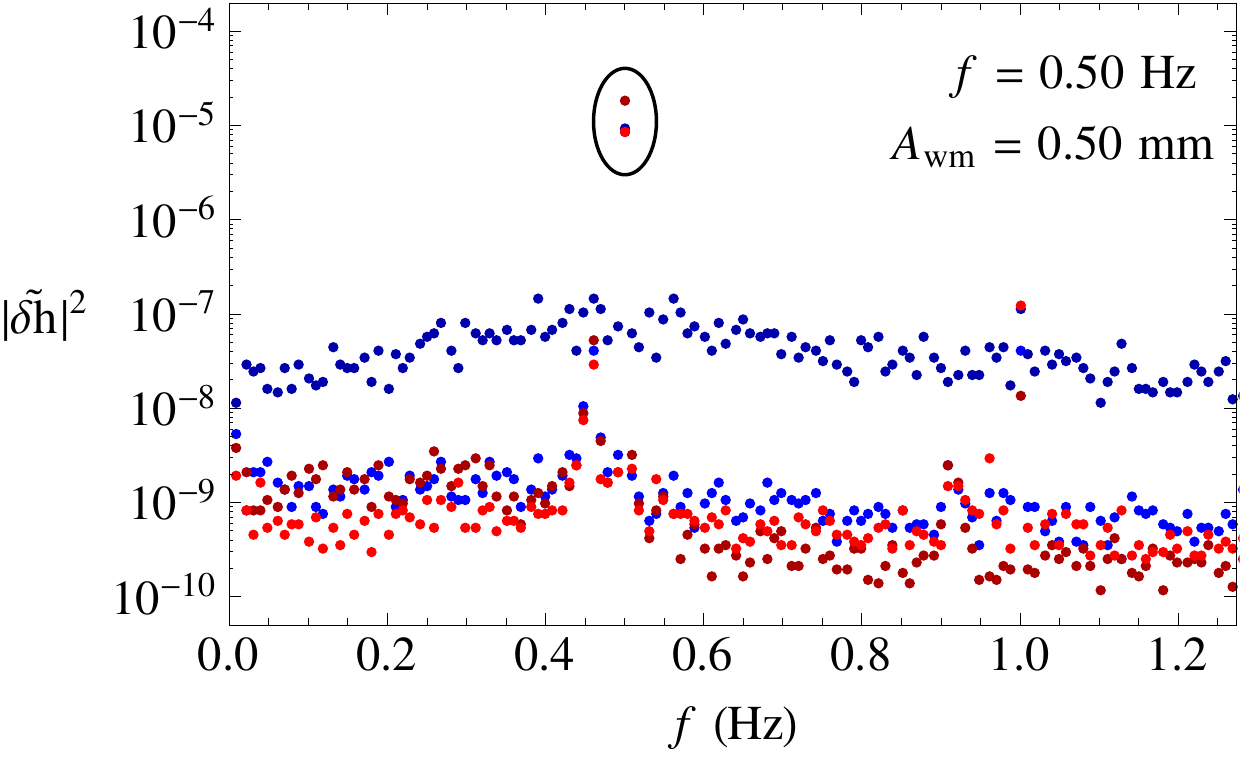} \\
\includegraphics[width=0.35\textwidth]{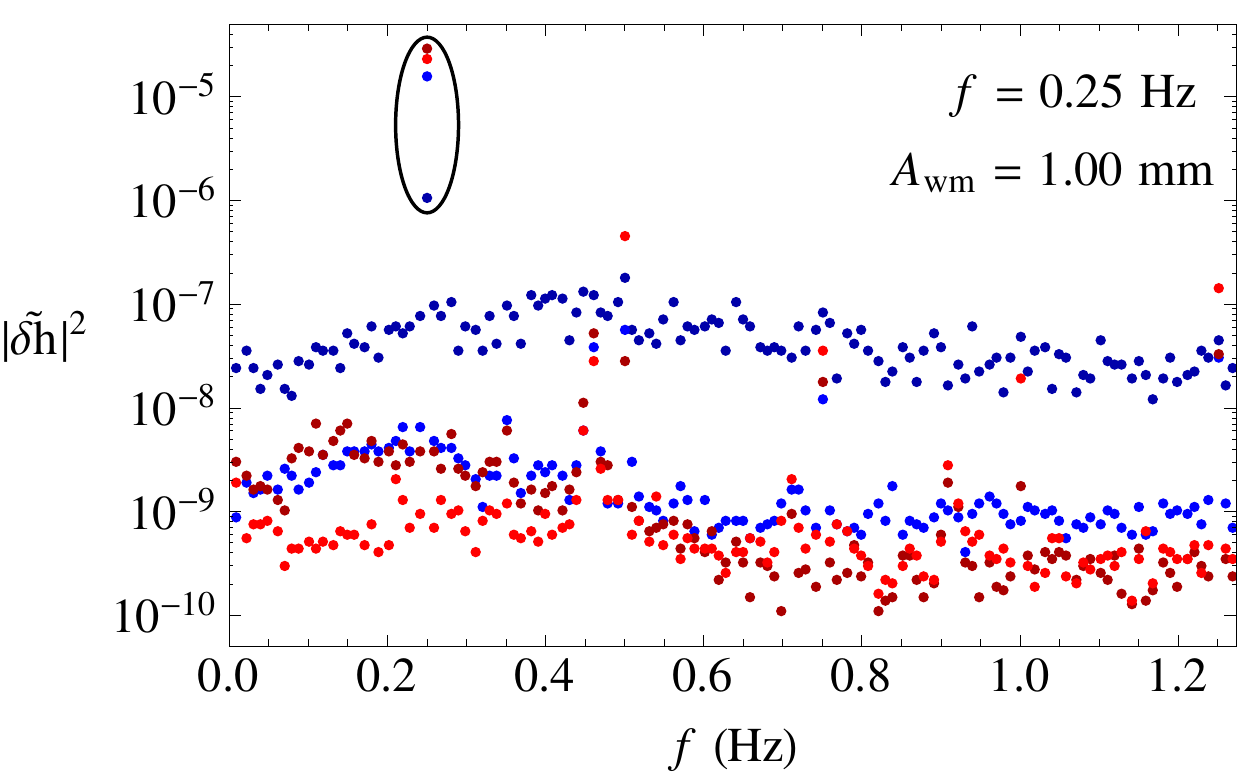} \, \includegraphics[width=0.35\textwidth]{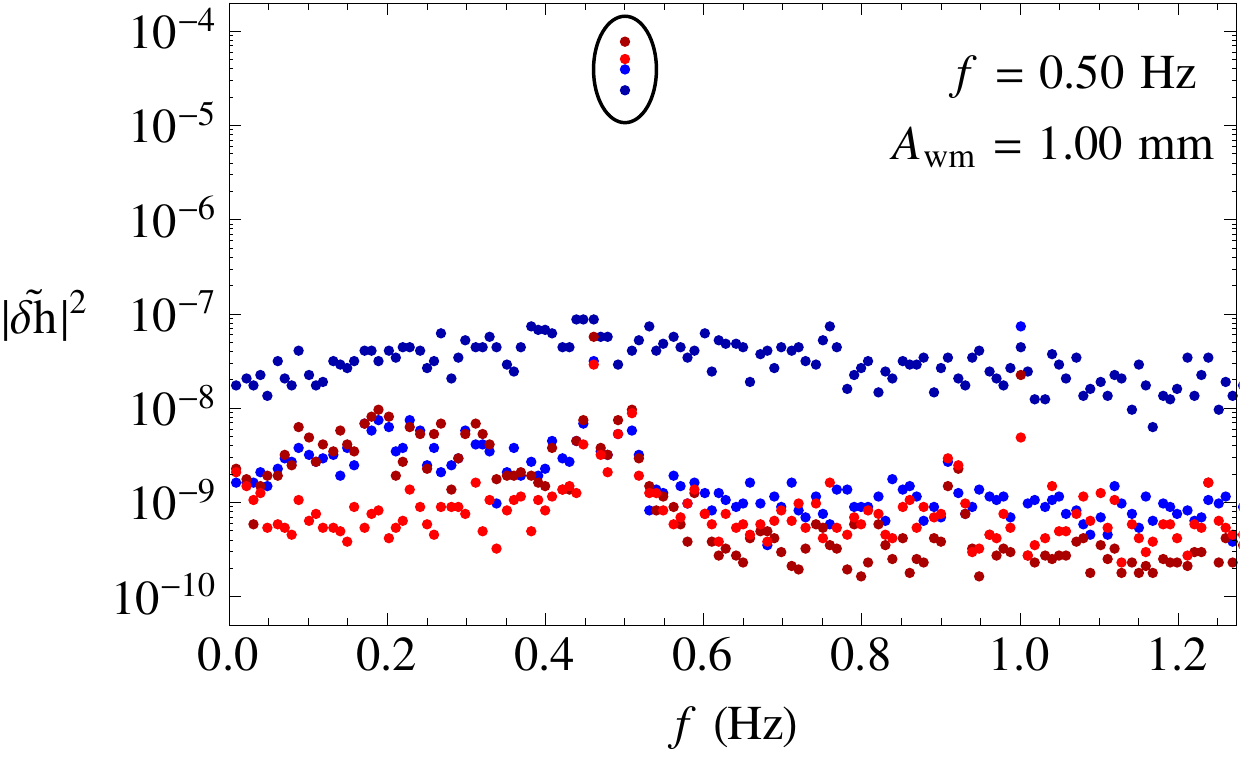}
\caption{Time-Fourier transform of the free surface deformations at several fixed positions. The different positions are indicated by the variously colored dots: light and dark red are, respectively, the upstream and downstream edges of the 
subcritical region (delineated 
by the dashed vertical 
lines in Fig.~\ref{fig:STdiagram_A050}); while light and dark blue are, respectively, the upstream and downstream edges of the 
supercritical region (delineated 
by the dotted vertical 
lines in Fig.~\ref{fig:STdiagram_A050}).  The left and right columns correspond to 
two different frequencies of the wave maker, while the upper and lower rows have 
two different amplitudes of the wave maker oscillations.  In each plot the main signal, whose frequency is equal to that of the wave maker, is highlighted by a black oval, and we note that its strength is always greater than the background noise by at least an order of magnitude.  Notably, while the other points have a comparable level of background noise, the downstream edge of the supercritical region shows noise which is about an order of magnitude greater.  This illustrates the growth in the noise level over the supercritical region already noted in Fig.~\ref{fig:STdiagram_A050}, and which may be due to the growth of side wakes (see Fig.~\ref{fig:SideWakes} below).  We also note that the pump frequency at $\sim 0.45\,\mathrm{Hz}$ can be seen in the background noise, except at the most downstream position where it has been obscured by the increased noise level.
\label{fig:FourXLog_A050}}
\end{figure*}

To investigate this further, in Figure~\ref{fig:FourXLog_A050} we show the squared magnitude of the time Fourier transforms of $\delta h(t,x)$, for several fixed values of $x$.  
To this end, the data is first divided into blocks of duration $100\,{\rm s}$ each, with the mean of $h(x)$ being calculated and subtracted in each block separately, and the squared magnitudes of the Fourier transforms are finally averaged over all eight blocks.
The chosen values of $x$ are taken to be the edges of the windows defining the near-homogeneous regions. 
There is some background noise of roughly constant magnitude over all frequencies, and which has roughly the same value in both the subcritical region and the upstream edge of the supercritical region.  However, by the time we reach the downstream edge of the supercritical region, this background noise has increased by between 1 and 2 orders of magnitude.  This seems to be consistent with the higher degree of noise seen in the supercritical region in Fig.~\ref{fig:STdiagram_A050}, and which emerges from within the near-homogeneous part of the supercritical region.

We also note the occurrence of a signal at a well-defined frequency $\omega \approx 2.9\,{\rm Hz}$ (or $f \approx 0.45\,{\rm Hz}$), which occurs at all positions except the downstream edge of the supercritical region, where it is drowned out by the increased background noise.  This frequency occurs for all stimulated frequencies and wave maker amplitudes, and we believe it to be characteristic of the vibrations of the water pump which maintains the background flow; indeed, this frequency shifts when the flow rate is changed, as would be expected if it were due to the water pump.  Interestingly, the measured values of $\left|R\right|^{2}$ and $\left|N\right|^{2}$ show signs of resonant behavior near $f = 0.45\,{\rm Hz}$, and this could be because the stimulated wave comes into resonance with the pump frequency, which could quite drastically alter the scattering process in the vicinity of that frequency.

Finally, we note that harmonics are visible in Fig.~\ref{fig:FourXLog_A050}, but these are quite clearly smaller than the main signal by several orders of magnitude.  We conclude that the signal is essentially linear in nature, and we can neglect the harmonics in predicting and explaining the observations.

\end{appendices}

\end{document}